\documentstyle[twocolumn,prb,epsfig,aps,floats]{revtex}
\begin{document}
\draft

\twocolumn[\hsize\textwidth\columnwidth\hsize\csname @twocolumnfalse\endcsname

\title{{\rm\small\hfill Phys. Rev. B, accepted (November 21, 2002).}\\
Sub-surface Oxygen and Surface Oxide Formation at Ag(111): \\ 
A Density-functional Theory Investigation}

\author{Wei-Xue Li$^{1}$, Catherine Stampfl$^{1,2}$ and 
Matthias Scheffler$^{1}$}
\address{$^{1}$Fritz-Haber-Institut der Max-Planck-Gesellschaft,
Faradayweg 4-6, D-14195 Berlin-Dahlem, Germany\\
$^{2}$Department of Physics and Astronomy, Northwestern University,
Evanston, Il. 60208-3112}
\maketitle

\begin{abstract}
To help provide insight into the remarkable catalytic behavior of 
the oxygen/silver system for heterogeneous oxidation reactions,
purely sub-surface oxygen, and structures involving
both on-surface and sub-surface oxygen,  as well as oxide-like
structures at the Ag(111) surface have been studied for a wide range of
coverages and adsorption sites using density-functional theory.
Adsorption on the surface in fcc sites is energetically favorable 
for low coverages,
while for higher coverage
a thin surface-oxide structure is energetically favorable. This structure
has been proposed to correspond to the experimentally
observed $(4\times4)$ phase.
With increasing O concentrations, thicker oxide-like
structures resembling compressed Ag$_{2}$O(111) surfaces are energetically
favored.
Due to the relatively low thermal stability of these structures,
and the very low sticking probability of O$_{2}$ at Ag(111),
their formation 
and observation
may require the use of atomic oxygen (or ozone, O$_{3}$) and
low temperatures.
We also investigate diffusion of O into
the sub-surface region at low coverage (0.11~ML), and the effect
of surface Ag vacancies in the adsorption of atomic oxygen and
ozone-like species.
The present studies, together with our earlier investigations of on-surface
and surface-substitutional adsorption, provide a comprehensive
picture of the behavior and chemical nature of the
interaction of oxygen and Ag(111),
as well as of the initial stages of oxide formation.
{\em Copyright (2003) by The American Physical Society.}
\end{abstract}

]

\section{Introduction}  

Investigation of the interaction 
of oxygen and silver, with regard to the role it plays in
technologically important heterogeneous catalytic reactions such as ethylene
epoxidation and the partial oxidization of methanol to formaldehyde,
has a long history.~\cite{sant87} Despite considerable efforts,
the O/Ag system is still not well understood on a microscopic
level, and the identification of active oxygen species involved
in the above-mentioned reactions, remains  unclear. 
Of particular interest is to understand the role and nature of
the elusive sub-surface O 
species in the function of silver as an oxidation catalyst.
Sub-surface oxygen
is thought to play a crucial role in the epoxidation of
ethylene (i.e., formation of ethylene oxide, alternatively known as epoxide), 
which takes place under atmospheric pressure and temperatures of 500-600~K,
and in the formation of formaldehyde from methanol, which can
occur either by
direct dehydrogenation or by oxi-dehydrogenation, and also takes place
under
atmospheric pressure and temperatures of 700-900~K 
and 800-900~K, respectively. 
\cite{bukh011,gran85,back80,felt80,wach78,rehren-91,segeth,bao-prb,herein-96,bao,nagy} 

Recent studies of oxidation reactions
over certain other metal surfaces have found that
metal oxides form and actuate the
reaction, in contrast to the hitherto believed pure metal,
i.e., for
the oxidation of carbon monoxide over Ru.~\cite{over,ss-500} 
One may therefore wonder if similarly silver oxides
form and play an important role.
The most stable oxide of silver, Ag$_{2}$O,
however, reportedly
thermally decomposes at 460 K and atmospheric pressure,
making this seem unlikely;
ruthenium dioxide on the other hand is stable up to 
temperatures greater than 1000~K.~\cite{here96}
To date there have been a limited number of {\em ab initio} studies of 
sub-surface oxygen adsorption and
oxide formation at transition metal surfaces.
These include the
investigation of oxygen incorporation into the Ru(0001) 
surface~\cite{stampfl-oprl,apa} which predicted that
after 1~ML of oxygen on the surface, O adsorbed in sub-surface sites
below the first Ru layer where an attractive interaction
between the oxygen atoms was identified, suggesting island formation.
Within these structures, the overall O-Ru bondstrength was reported to be 
significantly weaker than
pure on-surface adsorption, thus representing a significant destabilization
of the surface.
Subsequent investigations of this system in more detail
indicated that structures involving on-surface and
sub-surface oxygen functioned as
nucleation sites and metastable precursor structures
for bulk oxide formation and an atomic pathway for the transition
to RuO$_{2}$ was predicted.~\cite{karsten}
Similar studies for O/Rh(111)~\cite{piro}
have also recently been reported, as well
as studies of  the initial oxidation of alloy surfaces, 
i.e., NiAl(110).~\cite{finnis}
From a recent trend study of oxygen adsorption and incorporation
at the basal planes of Ru, Rh, Pd, and Ag, it was found that the
coverage at which the onset of
oxygen occupation of sub-surface sites occurs, correlates closely with
that predicted at which
bulk oxide formation begins, and that this occurs at progressively lower
coverages for the metals more to the right in the periodic table.~\cite{mira01}
The energy cost for lattice 
distortion was found to play a key role, where it is greater for 
the the metals more to the left  in the periodic table (i.e., Ru and Rh).
With regard to pure free-electron-like metals,
the O/Al(111)~\cite{kiejna}
and O/Mg(0001)~\cite{bungaro} systems have been investigated as well
by first-principles calculations.
Earlier studies using an embedded atom-type approach, focussed on 
the energy barriers for O penetration through the surface layer of
various metals and oxides,
which is an important consideration particularly with regard to
the kinetics.~\cite{nordlander}
Our present investigation
into the interaction of oxygen and Ag(111), in addition 
to sheding light on the understanding of how silver function as
an efficient oxidation catalyst,
also addresses the initial stages of oxide formation.  Our general
findings could 
have consequences for gold, also a 
noble metal oxidation catalyst~\cite{gold} 
which, for the (111) surface, exhibits a restructuring
and chemisorption of oxygen atoms at elevated temperatures
(500 -- 800~K) and
atmospheric pressure,~\cite{gold1,gold2}
very similar to Ag(111).
The findings of the
present study could also have consequences for copper, 
which catalyses the oxidation of methanol to formaldehyde.~\cite{copper}

Experimental
evidence indicates that the (111) orientation is an important crystal face
for real silver catalysts since at high temperatures,
facets  with this face result~\cite{bao,nagy} due to it
having the lowest surface energy.
There are only two ordered phases of oxygen on Ag(111) that have
been reported; 
the $(4\times4)$~\cite{rovi74} and 
$(\sqrt{3}\times\sqrt{3})R30^{\circ}$ structures.~\cite{bao-prb,bao} 
The latter actually 
exhibits a superstructure given in matrix
notation as ($26\times 1; -1\times 26)$. The $(4\times4)$ structure
has recently been investigated by  scanning tunneling microscopy (STM) 
where the atomic structure is proposed to involve a thin surface-oxide layer.
\cite{carl00,carl001} 
In this work (Ref.~\onlinecite{carl001}), it was furthermore proposed that at
low coverage ($\theta=0.05\pm0.03$~ML), O atoms occupied sub-surface
octahedral
sites below the first Ag layer. As will be shown in our present work, however, 
on-surface adsorption at low coverage is significantly more favorable.

The $(\sqrt{3}\times\sqrt{3})R30^{\circ}$ phase
has also been investigated by STM, as well as by
reflection electron microscopy (REM) and reflection high-energy electron
diffraction (RHEED). It forms only after exposure of 
silver (irrespective of the crystal face)  to O$_{2}$ at atmospheric
pressure and high temperatures (800-900~K).~\cite{bao-prb,bao} Various
atomic structures have been proposed for this phase; 
our recent STM simulation
supported a surface-substitutional geometry, however it 
was noted that this structure
was only {\em metastable}.~\cite{wxli01}  
To date, the atomic geometry of neither of
these ordered phases have been unambiguously confirmed
which 
calls for further quantitative structural analyses.

In our previous
work,~\cite{wxli01} we studied both on-surface oxygen and oxygen adsorption in
surface-substitutional sites on Ag(111) for a wide range of coverages. 
The bondstrength of
on-surface oxygen was found to be weak compared
to the transition metals to the left of silver in the periodic table
(e.g. Ru). 
With increasing coverage, there is a strong repulsion between on-surface oxygen
atoms, and on-surface oxygen adsorption becomes energetically
unstable with respect to gas phase molecular oxygen for coverages greater than
$\simeq$0.5~ML. 
In the present paper we investigate pure sub-surface oxygen and 
structures involving both oxygen adsorbed on the
surface and in the sub-surface region, hereafter denoted as 
``on-surface+sub-surface''. 
We also investigate
the proposed $(4\times 4)$ surface-oxide phase, and 
oxide-like structures containing
higher oxygen concentrations.
These results, together
with our earlier study, provide a comprehensive understanding of the
electronic and atomic structure, and stability of oxygen species
at the Ag(111) surface.
  
The paper is organized as follows: The calculation method is briefly explained
in Sec.~II, and in Sec.~III, results are presented for purely sub-surface O
structures (i.e. without on-surface oxygen) where the energetics, atomic
geometry and electronic properties for different sites and for a wide range of
coverages are described. Also energy barriers for oxygen penetration into the
sub-surface region are reported. 
In Sec.~IV structures are considered that
involve both on-surface and sub-surface oxygen species, as well
as the earlier proposed $(4\times 4)$ geometry.~\cite{carl00} 
In Sec.~V, geometries involving higher concentrations of oxygen
are considered, where the energetic preference of
oxide-like structures is found which have a geometry similar to the (111)
surface of Ag$_{2}$O. In this section, we also investigate an  ozone-like
species adsorbed at an Ag vacancy, as has been proposed to
be the active species for ethylene epoxidation.
The conclusions are given in Sec. VI. 

\section{Calculation Method \label{sec:cal}}

The density-functional theory (DFT) total energy calculations are
performed using the pseudopotential plane-wave method~\cite{fhi98} within the
generalized gradient approximation (GGA) for the exchange-correlation
functional.~\cite{pbe96,white94}  The pseudopotentials are generated by the
scheme of Troullier and Martins with the same functional.~\cite{fuch99,trou91}
The wave-functions are expanded in plane-waves with an energy cutoff of 50~Ry
and the surface is modeled by a five layer slab separated by 15 \AA\, of
vacuum. Oxygen is placed on one side of the slab where the induced
dipole moment is taken into account by applying a dipole
correction.~\cite{jorg92} The positions of the oxygen atoms and the top three
silver layers are relaxed until the forces on the atoms are less
than 0.015~eV/\AA\,. In the $(1\times1)$ Ag(111) surface unit cell, 21 special 
{\bf k}-points are used in the surface irreducible Brillouin zone (IBZ) for
the Brillouin-zone integration.~\cite{cunningham}  
These {\bf k}-points are equivalent for all of the surface structures
studied in order to maximize the accuracy when comparing
the energies of different coverages as calculated in different supercells.
We employ a
Fermi function with a temperature broadening parameter of
$T^{\rm el}=0.1$~eV, and the total energy is
extrapolated back
to zero temperature. Results of detailed convergence tests can be
found in Ref.~\onlinecite{wxli01}. 

The average adsorption energy per oxygen atom, $E_{\rm ad}$, 
is calculated according to,
\begin{equation}
E_{\rm ad}
=-\frac{1}{N_{\text {\rm O}}}[ E^{\text {\rm O/Ag(111)}}-
(E^{\text {\rm  Ag(111)}}+N_{\text {\rm O}}E^{\text {\rm free-O}})]
\quad ,
\label{eq1}
\end{equation}
where $N_{\rm O}$ is the total number of oxygen atoms
of the adsorbate-substrate system, and the total energy of
the adsorbate-substrate system, the clean surface, and the free oxygen atom
are represented by 
$E^{\rm O/Ag(111)}$, $E^{\rm Ag(111)}$, and $E^{\rm free-O}$, respectively. 
That is, $E^{\rm O/Ag(111)}$ represents the O/Ag system under
investigation, which e.g. may involve purely sub-surface O, or on-surface
and sub-surface species, or oxide-like structures.

To study the interaction between on-surface and sub-surface oxygen, we
consider how the stability of on-surface O is affected by sub-surface O (and
vice-versa). 
To do this, we define a so-called ``removal energy'', which is the energy
required to remove an on-surface O atom into the vacuum. It is given as,
\begin{equation}
  \label{eq:ebind}
E_{\rm on}^{\rm removal}
=-[E^{\rm O/Ag(111)}-(E^{\rm  O_{\rm sub}/Ag(111)}+E^{\rm free-O})] \quad ,
\end{equation}
where $E^{\rm O_{\rm sub}/Ag(111)}$ 
is the total energy of the ``reference'' system,
i.e., that containing only (one or more) sub-surface O species.  
$E_{\rm on}^{\rm removal}$ can also be thought of as the 
adsorption energy of an O atom onto the substrate which contains 
the sub-surface O atoms.
An obvious analogous equation holds for
the removal energy of a sub-surface O atom,
\begin{equation}
  \label{eq:ebind1}
E_{\rm sub}^{\rm removal}
=-[E^{\rm O/Ag(111)}-(E^{\rm  O_{\rm on}/Ag(111)}+E^{\rm free-O})] \quad ,
\end{equation}
where $E^{\rm O_{\rm on}/Ag(111)}$ 
is the total energy of the ``reference'' system,
i.e., that containing only the on-surface O species.  
We consider $E_{\rm sub}^{\rm removal}$ only for structures
having one O atom on the surface and one below it.
The adsorption and removal energies are
defined such that a positive number indicates that the adsorption is
exothermic (stable) with respect to a free O atom, and a negative number, 
endothermic (unstable). 

The definitions of quantities that we use to analyze our results, e.g.,
total and projected density of states (PDOS), density difference
distributions, and the work function, can be found in 
Refs.~\onlinecite{scheffler,wxli01}  

\section{Pure Sub-surface Oxygen \label{sec:sub}}

We first study sub-surface oxygen without the presence of on-surface
oxygen. For oxygen  occupation in the sub-surface region there two possible
sites: (i) the octahedral site, denoted hereafter as ``octa'', which has six
nearest-neighbour Ag atoms, three above and three below, and (ii) the
tetrahedral site, which has four nearest-neighbour Ag atoms. Due to the
presence of the surface, there are two types of tetrahedral sites; one is
where there are three silver atoms above it and one below, denoted as tetra-I,
and the alternative one, tetra-II, is just the opposite with one surface Ag
atom directly above, and three below it in the second Ag layer, as depicted in
the insets of Fig.~\ref{fig:e-sub}. 
We performed calculations for oxygen in these different
sites for a wide range of coverages, i.e., 0.11 to 1 ML. We focus on
adsorption immediately below the first Ag layer as we find that oxygen
adsorption deeper in the bulk is less favorable in every case (see Tab.~\ref{tab:sub}).
In the following subsections, the energetics, the atomic
and electronic structure, and energy barriers for O diffusion into
the substrate are discussed, respectively. 

\subsection{Energetics \label{sec:sub-en}}

\begin{figure}[t!]
\epsfxsize=0.45\textwidth \centerline{\epsfbox{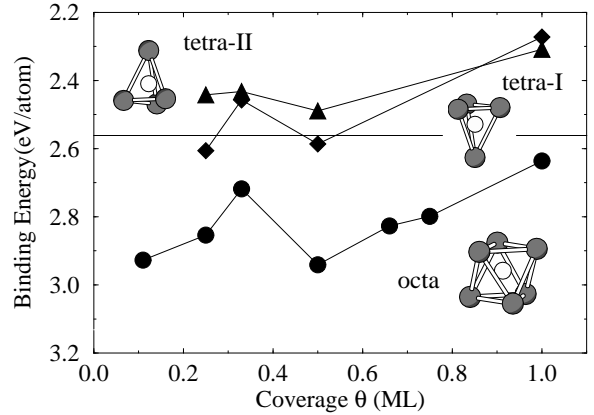}}
\caption{Calculated adsorption  energy of oxygen (with respect to
 atomic oxygen) under the first Ag(111) layer in octa
  (circles),  tetra-I (diamonds), and tetra-II  (triangles) sites, 
  as a function of coverage. 
  The horizontal full line represents half the experimental value of
  the binding energy of O$_2$ (2.56 eV).  
 The solid lines connecting the calculated adsorption energies are  to guide
  the eye.}
\label{fig:e-sub}
\end{figure}

The calculated adsorption energies, $E_{\rm ad}$, for O$_{\rm octa}$, 
O$_{\rm tetra-I}$, and O$_{\rm tetra-II}$, with respect to atomic 
oxygen, are     
plotted in Fig.~\ref{fig:e-sub} and are listed in Tab.~\ref{tab:sub}. 
For O$_{\rm octa}$ (circles),  
it can be seen  that the adsorption energy decreases when the coverage varies
from 0.11 to 0.33 ML, indicating a repulsive interaction between the 
O$_{\rm octa}$ atoms; however, it increases from 0.33 to 0.50 ML, so that the
effective interaction between the O$_{\rm octa}$ atoms  is
attractive. For coverages higher than 0.50~ML, the adsorption energy decreases
again. Despite this dependence on coverage, we note that the overall variation
in the magnitude of the adsorption energy is rather small, i.e. not greater
than 0.3~eV -- in sharp contrast to what we found for adsorption on the
surface, where the corresponding value is 1.93~eV.~\cite{wxli01} Compared to
O$_{\rm octa}$, the tetrahedral sites are energetically less favorable for the
whole coverage range.  
It can be noticed  that the tetrahedral sites exhibit a similar 
dependence on coverage as O$_{\rm octa}$. 

\begin{table}[t!]
\caption{
Calculated energies (in eV) and first metal interlayer expansion for various coverages 
of sub-surface oxygen in the octa, tetra-I,  and tetra-II sites. 
  $E_{\rm ad}$ is the adsorption energy with respect to atomic
  oxygen and $E_{\rm b}$ is the binding energy 
  on the pre-distorted surface, 
and $E_{\rm d}$ is the energy cost of the distortion
of the substrate as induced by sub-surface oxygen (without
the presence of O, see text).
  $\Delta d_{\rm 12}$ is the percentage change in the first
  metal interlayer distance, where the center-of-mass of each layer is
  used. The calculated bulk interlayer distance is 2.43 \AA\,. Where available,
values of the adsorption energy
are given for O occupation under the second metal layer,
$E_{\rm ad}$(2nd).}
\label{tab:sub}
\begin{tabular}{cc|ccccccc}
 \multicolumn{2}{c} {$\theta$(ML)}
                & 0.11 & 0.25 & 0.33 & 0.50 
                & 0.66 & 0.75 & 1.00 \\  \hline
 \multicolumn{2}{l}{octa}
                &      &      &      &
                &      &      &    \\ \hline
 & $E_{\rm ad}$
                &2.93   &2.86   &2.72   &2.94  
                &2.83   &2.80   &2.64     \\ 
 & $E_{\rm b}$     
                & -     &3.40   &3.16   &3.40  
                & -     &3.15   &2.91     \\ 
 & $E_{\rm d}$    
                & -   &0.54   & 0.44   &0.46  
                & -   &0.35   & 0.27     \\ 
 & $\Delta d_{\rm 12}$     
                & 6.4   & 10.9   &  10.5  & 21.2  
                & 28.6  & 29.9   &  33.2     \\ \hline
 & $E_{\rm ad} ({\rm 2nd})$ & 2.72  & 2.64  &2.70&-&-&-&- \\
 \hline
 \multicolumn{2}{l}{tetra-I}
                &      &      &      &
                &      &      &    \\ \hline
 & $E_{\rm ad}$
                &-      &2.61   &2.46   &2.59  
                &-      &2.45   &2.27     \\ 
 & $E_{\rm b}$     
                & -     &3.49   &3.14   & 3.32  
                & -     &2.97   &2.65    \\ 
 & $E_{\rm d}$    
                &-     & 0.88   &  0.68  & 0.73  
                &-     & 0.52   &  0.38   \\ 
 & $\Delta d_{\rm 12}$     
                & -     & 15.5   &  14.2  & 27.4  
                & -     & 41.1   &  46.8     \\ \hline
 & $E_{\rm ad} ({\rm 2nd})$ &2.31    &2.19 &-&-&-&-&-\\
 \hline
 \multicolumn{2}{l}{tetra-II}
                &      &      &      &
                &      &      &    \\ \hline
 & $E_{\rm ad}$
                &-     &2.44   &2.43   &2.49  
                &-     &-      &2.31      \\ 
 & $\Delta d_{\rm 12}$     
                & -      & 17.5   &  20.4  & 37.3  
                & -     & -      &  49.2      \\ 
\end{tabular}
\end{table}
From our calculations the {\em relative stability} of
the various systems are well defined, the absolute value of the adsorption
energies is less so due to the error of standard DFT calculations in describing
the free O atom and molecule.~\cite{wxli01} In particular, the experimental
binding energy of O$_{2}$ per O atom is 2.56~eV while the calculated value is
3.16 
eV/atom,~\cite{wxli01} that is, the theoretical value is significantly
overestimated. The deviation to the experimental value highlights the
difficulty in  obtaining quantitatively, the absolute adsorption
energies. This is particularly problematic for this system as can be
appreciated from Fig.~\ref{fig:e-sub}:  On consideration of the
theoretical value of 
3.16~eV,
it would indicate that all of the sub-surface sites
are unstable with respect to the oxygen molecule, while the experimental value
indicates that the octa site is stable with respect
to O$_{2}$ for all the coverage range. The
adsorption energy of oxygen on the substrate is expected also to be overbound, 
although not nearly so severely as the O$_{2}$ molecule.

Compared to sub-surface adsorption, on-surface 
adsorption is notably more favorable for low coverages: compare
3.61~eV~\cite{wxli01} (fcc) to 2.93~eV (octa, Tab.~I) for coverage 0.11~ML and
3.52~eV~\cite{wxli01} (fcc)  to 2.85~eV (octa, Tab.~I) for coverage 0.25~ML.
Due to the build-up of a strong repulsion between on-surface
adsorbates,~\cite{wxli01} for a coverage of 0.50~ML the energies of 
on-surface and sub-surface oxygen become practically degenerate (compare
2.92~eV~\cite{wxli01} (fcc) and  2.94~eV (octa, Tab.~I)). For
higher coverages, on-surface adsorption becomes considerably less favorable
than  sub-surface oxygen

\begin{figure}[t!]
\epsfxsize=0.45\textwidth \centerline{\epsfbox{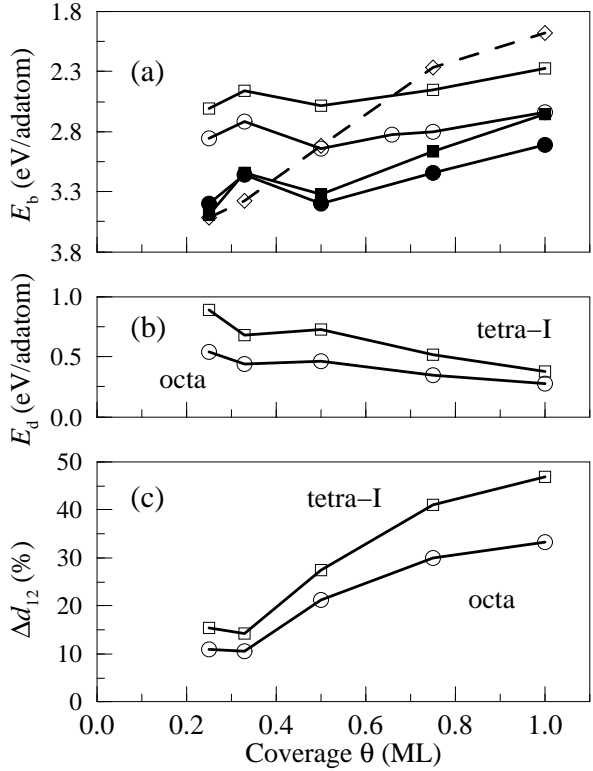}}
\caption{(a) Binding energies of O on the pre-distorted substrate, 
$E_{\rm b}$,
  in the octa (filled circles) and tetra-I sites (filled squares)
and the corresponding adsorption energies, $E_{\rm ad}$, of O in the
octa (open circles) and tetra-I (open squares) sites.
The adsorption energies of  on-surface fcc-O are
also shown (open diamonds). 
  (b)  Energy cost, $E_{\rm d}$,  of distorting the substrate
  lattice to the geometry induced by sub-surface oxygen (but without the
  presence of oxygen) for the octa (open circles) and tetra-I (open squares)
sites. 
(c) Percentage 
  change of the first metal interlayer distance
  compared to the bulk value for the octa (circles) and tetra-I (squares)
  sites.}
\label{fig:e-sub-decomp}
\end{figure}

The electron density at sub-surface sites is significantly greater 
than for on-surface sites
which gives rise to a strong kinetic repulsion due to
the orthogonalization of the wavefunctions of the substrate caused by the
presence of oxygen. To reduce this repulsion, the lattice expands to an
optimum value.~\cite{chak85} In order to aid our discussions concerning
on-surface versus sub-surface adsorption, it is constructive to regard the
sub-surface adsorption energy, $E_{\rm ad}$,  as consisting of two
contributions: (i) the energy cost of distorting the (clean) substrate lattice
to the geometry induced by sub-surface oxygen, $E_{\rm d}$, and (ii) the
binding energy of oxygen on the pre-distorted substrate, 
$E_{\rm b}$, i.e., 
$E_{\rm ad} = E_{\rm b} - E_{\rm d}$. We may then regard the 
binding energy as
being due only to electronic effects, as we have removed the energy cost of
distorting the lattice; that is, as  purely reflecting the bond strength of 
the O-Ag interaction. 
We have calculated these contributions for
the octa and tetra-I sites which 
are given in Tab.~\ref{tab:sub} and are also plotted
in Fig.~\ref{fig:e-sub-decomp}, along with the corresponding on-surface
adsorption energies~\cite{wxli01} for comparison. In addition,
the percentage change
in the first metal-metal interlayer distance is shown.

Comparing the binding energies of O on the pre-distorted surface
with on-surface adsorption (see Fig.~\ref{fig:e-sub-decomp}a), 
it can be seen that for coverages 0.25 and 0.33~ML,
on-surface adsorption is still (slightly) energetically more favorable
than sub-surface adsorption in either the octa or tetra-I sites, thus the
preference for O to be on the surface is not due to the energy cost of
distortion of the substrate for the case of sub-surface adsorption. It can be
seen, however (from e.g. Fig.~\ref{fig:e-sub-decomp}b) that the energy cost of
distortion is significant and does reduce the adsorption energy of
sub-surface oxygen substantially. For example, at 0.50~ML the binding energy of 
sub-surface oxygen
in the octa site on the pre-distorted substrate is 0.48~eV greater
than adsorption on the surface, but the
{\em adsorption} energy (i.e. when paying for the distortion cost)
is practically the same. 
Furthermore, it can be seen that 
like the adsorption energy, only to a greater degree,
the binding energy of sub-surface
oxygen decreases in the coverage range 0.50 - 1.0~ML.
This repulsive interaction
is attributed to unfavorable O-O interactions due to the close proximity
of the partially negatively charged
O atoms to one another and the larger expansion of the metal-metal
spacing for these coverages, which less effectively screens the O atoms.

The energy cost of distorting the substrate (Fig.~\ref{fig:e-sub-decomp}b) 
per O atom generally decreases with increasing O coverage due to 
an effective sharing of the 
energy cost by neighboring O atoms. The exception,
from coverage 0.33 to 0.50~ML,
where the distortion energy slightly increases,
is because there is a significant increase of the metal interlayer
spacing for the latter coverage 
(see Fig.~\ref{fig:e-sub-decomp}c).
The distortion energy is always larger for the tetra-I 
site, which is understandable due to the 
larger oxygen-induced expansion of the first metal 
interlayer spacing (see Tab.~\ref{tab:sub} and Fig.~\ref{fig:e-sub-decomp}c)
caused by the smaller space in
which to accommodate the sub-surface O atom. 

As seen clearly from
Fig.~\ref{fig:e-sub}, the adsorption energy
for the octa site is greater than
the tetrahedral sites for all the coverages considered.
If we remove, however, the distortion energy, we
see from Tab.~\ref{tab:sub}  (and Fig.~\ref{fig:e-sub-decomp}a) that for
coverage 0.25~ML, the tetra-I site has in fact slightly
greater binding energy (compare 3.49 to 3.40~eV). In this case it
is the larger cost of distortion for the tetra-I site
(compare 0.88 to 0.54~eV)  that results in the octa
site having the larger adsorption energy. For all
the higher coverages, however, the octa site has the stronger O-Ag binding    
and even though the energy cost of distortion for the tetra-I site is larger
than the octa site in all cases, 
this does not determine the adsorption site
preference and make the tetra-I site less favorable than the
octa site. The reason that the
binding energy of the tetra-I site is less favorable than the octa site for
coverages greater than 0.25~ML, is therefore an electronic effect. 
This could be due to a less efficient screening by the metal
atoms of the partially negatively charged sub-surface O atoms,
due to the larger Ag-Ag interlayer spacing of the tetra-I site as
shown clearly in Fig.~\ref{fig:e-sub-decomp}c. 

The above findings highlight the interplay between
the energy gain due to bond formation and the energy cost due to structural
deformation. These
effects are materials dependent: To the left of Ag in the periodic table,
the transition metals are ``harder'' (i.e.,
have higher bulk moduli), and the energy cost of deformation is 
higher;~\cite{mira01}
on the other hand, the energy gain due to bond formation between oxygen and
the transition metal is stronger as well.

\begin{figure}[t!]
\epsfxsize=0.35\textwidth \centerline{\epsfbox{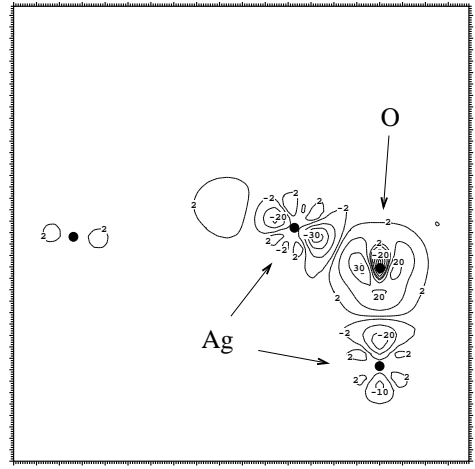}}
\caption{The difference electron density, $n({\bf r})^{\Delta}$, 
(see the definition in Ref.~\protect\onlinecite{wxli01})
 for oxygen in the sub-surface tetra-I site for coverage 0.25~ML. The  
  contour plane is in the [211] direction and is perpendicular to 
  the
  Ag(111) surface. The unit is  ${\rm 10^{-3}~Bohr^{-3}}$. 
  The positions of the O and Ag atoms are indicated by the arrows.
}
\label{fig:tetra-I}
\end{figure}

\subsection{Atomic structure and electronic properties\label{sec:sub-st}}

As seen above, sub-surface oxygen occupation induces 
a significant structural distortion of the Ag(111) substrate as shown by 
the 
variation of the first interlayer spacing in Tab.~\ref{tab:sub} and in 
Fig.~\ref{fig:e-sub-decomp}c: Namely, 
the expansion varies from 
0.16 to 0.81~\AA\,
for O$_{\rm octa}$ when $\theta$ increases from 0.11 to 1.00~ML, and
from 0.38 to 1.14~\AA\, for O$_{\rm tetra-I}$, and from 0.43 to 1.20~\AA\ for
O$_{\rm tetra-II}$ when $\theta$ increases from 0.25 to 1.00~ML. This is in 
sharp contrast to what we found in our study of on-surface oxygen adsorption
where the contraction of the first interlayer spacing induced by on-surface
oxygen was less than 0.02~\AA\, for the whole coverage range investigated
(0.11 to 1.00~ML).~\cite{wxli01}  
The large expansion is,
however, similar to what has been found for sub-surface O at 
other metal surfaces.

The O-Ag bondlengths are generally longer for the case of the
octa site compared to the tetra sites, which is
in-line with the understanding that adsorbates in higher coordinated sites
have longer bondlengths compared to lower coordination sites.~\cite{scheffler,pauling}
Most of the O-Ag bondlengths involving
sub-surface oxygen are longer than the
typical bondlength between on-surface oxygen and Ag (2.17~\AA\,),
i.e., roughly 8.8, 2.3, 2.8\% longer for the octa, tetra-I, and tetra-II  
sites, respectively. 

\begin{figure}[t!]
\epsfxsize=0.45\textwidth \centerline{\epsfbox{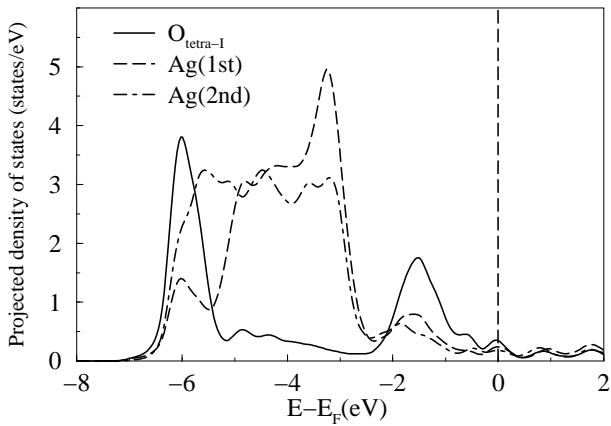}}
\caption{Total projected density of states (PDOS) for 
  oxygen in 
  the sub-surface tetra-I site for
  coverage 0.25~ML.
  The Ag PDOS of the first and second metal layer atoms
  that are bonded
  to oxygen are indicated by  dashed and dot-dashed lines,
  respectively, and the O PDOS 
  is indicated by a solid line. 
  The vertical dashed line marks the Fermi energy.
\label{fig:dos-sub}}
\end{figure}

The change in the work function induced by sub-surface oxygen is modest, i.e., 
less than 0.56~eV for all of the adsorption sites, throughout the range of
coverages considered, which is in strong contrast to the case of on-surface
oxygen adsorption, where it is as large as 4~eV at 1.00~ML.~\cite{wxli01} 
This is
due to the location of O under the surface Ag layer, which
screens well the partially negatively charged oxygen atoms, 
and in addition there is, to a large degree, a
cancellation of dipole moments involving oxygen and the top two Ag
layers as indicated by the difference electron density in Fig.~\ref{fig:tetra-I}.

The electron density of the exposed (surface) silver atoms and the silver
atoms bonded to O in the second layer, is depleted,
while there is a large accumulation on the O atom.
The result is very similar for the octa site.

Similarly to on-surface oxygen adsorption,~\cite{wxli01} the interaction
between sub-surface oxygen and silver is mainly via hybridization of 
the O-2$p$ and Ag-4$d$-5$sp$ states, where antibonding states are largely 
occupied, indicating an ionic nature and explaining the relatively weak 
binding energy compared to the metals to the left of Ag in the periodic 
table. This can be seen from the total projected density of states of the 
O atom 
and of the Ag atoms in the first and second layers that are bonded to O, as 
shown in Fig.~\ref{fig:dos-sub} for oxygen in the tetra-I site at
coverage 0.25~ML. The band width of the Ag-4$d$ states are  notably narrower 
for the surface Ag atoms due to the lower coordination. 
As for on-surface oxygen adsorption, no other peaks occur between 
the O-2$s$ state and the lower edge of the Ag-4$d$ band; we point this out 
because such a feature has been found in some experimental studies, the origin 
of which is under debate (see e.g., Ref.~\onlinecite{bukh011}). Clearly, purely 
sub-surface oxygen is not responsible for these features, nor is purely on-
surface
oxygen or surface-substitutional oxygen.~\cite{wxli01}
This will be discussed in more detail in Sec.~VB.
Compared to on-surface oxygen, the O-2$s$ PDOS for sub-surface oxygen is at a
{\em lower} energy, which,
from the initial state theory of core-level shifts,
indicates that sub-surface oxygen is less negatively
charged.

\subsection{Energy barriers for O penetration into the substrate
\label{sec:sub-kin}}

\begin{figure}[b!]
\epsfxsize=0.45\textwidth \centerline{\epsfbox{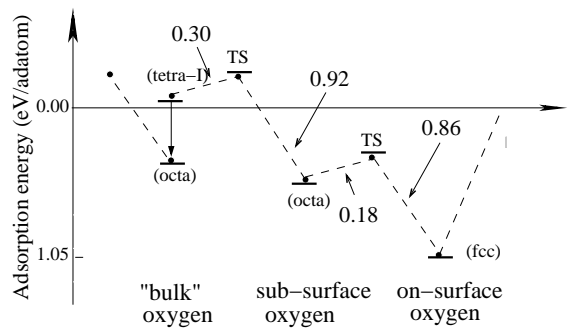}}
\caption{Adsorption energy of oxygen at a coverage of 0.11~ML
for penetration into the sub-surface octa 
site from the on-surface fcc-hollow site through the first
Ag layer, and from the octa site to the ``bulk'' tetrahedral site,
through the second Ag layer (from right to left). From the tetra-I site
O will move to the more favorable ``bulk'' octa site under the second
Ag layer.
The short lines indicate the adsorption energy with
respect to half of the binding energy of the oxygen molecule (experimental
value). ``TS'' represents the transition states, 
and the numbers on the arrows indicate the energy barriers.
}
\label{fig:poten}
\end{figure}

To investigate sub-surface oxygen in more detail with regard to the 
energetics of its formation, we study the penetration of O through
the surface Ag layer from the on-surface (fcc) site to the sub-surface 
octahedral site at low coverage (0.11~ML). The transition state corresponds to 
the oxygen atom in the plane of the surface Ag(111) layer.
We fully relax the first three Ag layers, but to stay at the transition 
state, we restrain O to be in the plane of the first Ag(111) layer 
(if we did not do this, the O atom would move to one of the two minima 
either side). The vertical distance of this top layer, however, is relaxed.
We also considered the diffusion barriers for oxygen under the first
Ag layer to move deeper into the bulk, i.e. to diffuse through the
second Ag layer by analogous calculations. In this case the pathway is from
the octa site under the first Ag layer to the tetra-I site under the second 
Ag layer. Oxygen will then
move to the energetically more favorable octa site
under the second Ag layer.
We note that due to the use of a $(3\times3)$ surface cell, which will
restrict complete lateral relaxations at the transition state,
the barriers could be slightly lower. The present values may be
taken as an upper limit.
From Fig.~\ref{fig:poten} it can be seen that the energy
barrier for fcc-oxygen to enter the sub-surface region is 0.86~eV.
For the reverse process, i.e., to diffuse to the
surface from the octa site, the barrier is only 0.18~eV.
Assuming that the prefactor for the two diffusion
directions is the same, and using a simple Arrhenius equation, the ratio of the 
diffusion coefficients for movement from the surface to the sub-surface,
compared to from the sub-surface to the surface is: 5.1$\times10^{-8}$ at 
470~K and 1.6$\times10^{-4}$ at 900~K. These kinetic considerations
indicate that chemisorbed oxygen will stay on the surface at low
coverage and at temperatures of around 470~K. Even at the higher temperature 
of 900~K, the difference is still significant.

The obtained energy barrier to move deeper into the bulk from under the first 
Ag layer 
to under the second Ag layer
is 0.92~eV.
This  value is slightly 
larger compared to the case of oxygen diffusing through the first metal layer; 
this is due to the presence of the surface in that it costs less energy for the 
surface Ag atoms to relax compared to the Ag atoms in the second layer, which 
are 
more constricted due to other Ag atoms both above and below. 
Although configurational 
entropy for O in the bulk could drive oxygen to diffuse deeper 
into bulk region, the above results clearly show
that there is large potential gradient indicating
that the oxygen, in the lower coverage regime,
will be concentrated near surface.

As noted in the introduction,
a chemisorbed oxygen species at coverage 0.05$\pm$0.03~ML was observed,
by  scanning tunneling microscopy (STM).~\cite{carl001}
The associated feature exhibited a depression with a diameter of
9$\pm$1\AA\,  which, according to the STM stimulation,
suggested that the oxygen atom is located in the octa site under
the first silver layer. On-surface adsorption was excluded
since the diameter of the depression in the stimulations was
significantly less, i.e. just 4$\pm$1\AA\,.
The tetra-I site was excluded on the basis that it offered
less space and was expected to be energetically unfavorable. 
The alternative tetra-II site was not mentioned.
No calculations, however, of energies were performed in this study.
As our results show, at coverage 0.11~ML, the binding energy of O 
on the surface
is energetically significantly more favorable than sub-surface oxygen
which does not support this assignment. 
The effective O-O repulsion is 
not very strong in the coverage range 0.11 to 0.25, i.e. the adsorption
energy is only 0.09~eV smaller for the latter coverage.
Thus we believe that for the lower coverage of 0.05 ML, the
energetic preference for on-surface adsorption will not change.

\section{ Effect of Sub-surface Oxygen \label{sec:eff}}

In the previous section it was shown how 
on-surface oxygen was energetically favorable for coverages less
than $\approx$0.50~ML compared to pure sub-surface adsorption, but for higher
coverages, sub-surface adsorption was preferred.
It was also seen how the presence of sub-surface
oxygen changes the nature of the surface Ag atoms, namely, by causing
a depletion of the electron density.
In order to investigate the effect that sub-surface oxygen has on the 
reactivity of the surface, as a first  
step we study the adsorption of additional oxygen
on the surface with sub-surface O present. 
This study is interesting in the sense that if
on-surface oxygen is taken to represent
an arbitrary electronegative adparticle,  
its behavior may reflect a general tendency.
Investigating the interaction between on- and sub-surface oxygen
is also essential in relation to understanding the initial formation of
surface oxides. 
We first consider many different atomic arrangements involving
structures with an O atom on the surface and one under
the surface Ag layer in order to identify energetically
favorable geometries.

\subsection{Energetics \label{sec:eff-en}}

\begin{figure}[t!]
\epsfxsize=0.45\textwidth \centerline{\epsfbox{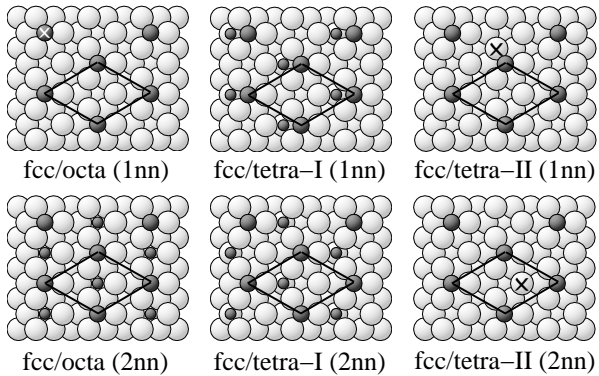}}
\caption{
Schematic geometry of 
the $(2\times2)$ surface unit cells with
on-surface oxygen in the fcc site, O$_{\rm fcc}$, (larger dark circles)
for a coverage of 0.25~ML. The various sub-surface
sites are depicted as the small dark circles and the 
resulting on-surface+sub-surface
geometries defined by the label (e.g. fcc/octa (1nn)).
There are two different types of sub-surface sites;
ones that are closest to O$_{\rm fcc}$,
denoted `1nn' and ones that are further away, denoted
`2nn'.
Note, as indicated by the `{\sf x}' 
on the atoms, the 1nn-O$_{\rm octa}$ site is located
directly below O$_{\rm fcc}$ and the tetra-II site is located directly below surface
Ag atoms.
}
\label{fig:schem}
\end{figure}

We consider the case of one on-surface oxygen atom in
the fcc site, O$_{\rm fcc}$,  of the $(2\times2)$ cell plus
one sub-surface oxygen atom, i.e., a total coverage of 0.50~ML. 
With this arrangement,
there are three high-symmetry sub-surface sites,
namely, octa, tetra-I, and tetra-II, 
as described above, but due to the presence of 
O$_{\rm fcc}$, there are now {\em two} different 
kinds of each of these sites, which can be differentiated
by their relative distance to O$_{\rm fcc}$.
We  therefore assign the label ``1nn'' (1st nearest-neighbor) 
to indicate the three sub-surface sites closest to O$_{\rm fcc}$ and
``2nn'' to indicate those that are the
second nearest neighbor sub-surface sites.
These six possible sub-surface sites are shown in Fig.~\ref{fig:schem}.
We tested all of these possibilities, where the
positions of the oxygen atoms and top three silver metal layers were
fully relaxed.  Similar geometries
exist when on-surface oxygen is located in the hcp-hollow
site; in the light of what is learnt from the results
for O in the fcc site however, we viewed
it necessary to only consider two structures for this case. 

We first discuss results for sub-surface
adsorption with on-surface O in fcc sites. The
calculated (average) adsorption energies are given in 
Tab.~\ref{tab:sub1}, along with the removal  energies (cf. Eqs.~2 and 3),
and the O-Ag bondlengths and metal interlayer
expansions, as well as the work function change.

It can be seen that the values depend strongly
on the location of sub-surface oxygen: The average
adsorption energies vary from 2.77 to 3.27~eV. 
Furthermore, the 1nn set of structures 
are energetically unfavorable compared to the 2nn set. 
This is due to the close proximity of the former to O$_{\rm fcc}$ 
which
causes a considerable O-O repulsion.
For example, the adsorption
energy of the  {\em most favorable}
structure of the 1nn set  
is 2.92~eV, while the {\em least
favorable} structure of the 2nn set is 2.99~eV. 
The energetically most favorable structure of all is that where sub-surface
oxygen is located in the tetra-I site of the 2nn set. 
This structure is
denoted as $({\rm O}_{\rm fcc}/{\rm O}_{\rm tetra-I})_{\theta=0.50}$  
and can be seen more clearly in Fig.~\ref{fig:oxide}b.

To investigate the affect of sub-surface oxygen on the bond strength
of O$_{\rm fcc}$, we 
consider the removal energy cf. Eq.~2.
We calculated this quantity for
O in all of the various sub-surface sites.
For sub-surface O in the octa, tetra-I, and tetra-II
sites of the 1nn set, the respective removal energies 
are: 2.98, 2.94 and 3.36~eV.  For the same sites but in the
2nn set, the values are 3.10,
3.93 and 3.64~eV, respectively. These values can be compared to
the adsorption  energy of O$_{\rm fcc}$  
(at 0.25~ML)
without the presence of sub-surface oxygen  which
is 3.52~eV.~\cite{wxli01} 
Thus,
depending on the location of sub-surface oxygen, it can either 
{\em destabilize or stabilize} significantly the interaction between on-surface
oxygen and the metal substrate.
The energetically most stable structure, 
$({\rm O}_{\rm fcc}/{\rm O}_{\rm tetra-I})_{\theta=0.50}$,
{\em stabilizes} on-surface O by 0.41~eV (compare 3.93 to 3.52~eV). 
The affect of the presence of on-surface O on the binding
of sub-surface O (cf. Eq.~3)
is like that found for the affect of sub-surface
O on the binding of on-surface O. This can be seen from inspection
of the values of $E_{\rm sub}^{\rm removal}$ in Tab.~II,
together with the values of $E_{\rm ad}$ in Tab.~I for structures
involving purely sub-surface oxygen.

\begin{table}[t!]
\caption{Calculated atomic 
geometries, energetics and work function change 
for oxygen adsorption in structures involving both on-surface 
fcc-hollow sites and sub-surface sites at a total coverage of 0.50~ML. 
$R_{\rm 1}^{\rm on}$ is the bondlength between 
on-surface oxygen and silver atoms in the first metal layer, while 
$R_{\rm 1}^{\rm sub}$   
is the bondlengths between sub-surface oxygen and silver atoms in the first
metal layers. $d_{\rm 12}$ is the first 
metal interlayer distance given with respect to the
center of gravity of each layer.
  The bulk interlayer distance is 2.43 \AA\,.
  $E_{\rm ad}$ is the average adsorption energy
  with respect to atomic oxygen, while 
  $E_{\rm on}^{\rm removal}$ and $E_{\rm sub}^{\rm removal}$ 
are the removal energies for on-surface and
sub-surface oxygen atoms, as defined in Eqs.~\ref{eq:ebind} and \ref{eq:ebind1}.  
$\Delta\Phi$ is the change in work function with
respect to the clean surface (calculated to be 4.45~eV). 
The unit of length is \AA \, and energy is
  eV. }
\label{tab:sub1}
\begin{tabular}{l|llllllll}
     &$R_{\rm 1}^{\rm on}$ &$R_{\rm 1}^{\rm sub}$ 
     &$d_{\rm 12}$ &$E_{\rm ad}$&$E_{\rm on}^{\rm removal}$&
$E_{\rm sub}^{\rm removal}$ & $\Delta\Phi$ \\ \hline
1{\rm nn}          &      &      &      &     &    &    &   \\ 
fcc/octa      &2.17  &2.46    &2.73  &2.92 &2.98&2.33&1.81 \\  
fcc/tetra-I   &2.19  &2.18   &2.81  &2.77 &2.94&2.03&1.52 \\  
fcc/tetra-II  &2.18  &2.26    &2.80  &2.90 &3.36&2.29&1.10 \\  \hline
2{\rm nn}          &      &     &       &      &     &    &    &   \\ 
fcc/octa      &2.19  &2.26    &2.72  &2.99 &3.10&2.44&1.40 \\  
{\bf fcc/tetra-I} &{\bf 2.10}  &{\bf 2.09}   &{\bf 2.89}  
              &{\bf 3.27} &{\bf 3.93}&{\bf 3.02}&{\bf 1.15} \\  
fcc/tetra-II  &2.21  &2.17   &2.84  &3.05 &3.64&2.58&0.84 \\  
\hline
{\bf hcp/octa} &{\bf 2.15} &{\bf 2.11} &{\bf 2.78}  
              &{\bf 3.33} &{\bf 3.80}  &{\bf 3.25} &{\bf 1.02} \\  
hcp/tetra-I   &2.16  &2.18  &2.86  &2.77 &2.94&2.14&1.59 \\  
\end{tabular}
\end{table}

For on-surface oxygen in the hcp-hollow site, we considered the 
octa and the tetra-I sites of the 2nn set for sub-surface O.
The former, denoted as
$({\rm O}_{\rm hcp}/{\rm O}_{\rm octa})_{\theta=0.50}$, 
is geometrically very similar to the energetically favorable structure,
$({\rm O}_{\rm fcc}/{\rm O}_{\rm tetra-I})_{\theta=0.50}$,
namely, both involve a linear O-Ag-O
bonding arrangement.  And as can be seen from 
the results presented in Tab.~\ref{tab:sub1}, it is 
also
a low energy  structure; in fact, it is the lowest of all. That it is slightly
more favorable (by 0.06~eV) 
may be understood due to the fact that the pure sub-surface O$_{\rm octa}$
structure at 0.25~ML is energetically
more favorable than the O$_{\rm tetra-I}$ site (by 0.24~eV)
as shown in Tab.\ref{tab:sub}, while the difference between
pure on-surface O$_{\rm fcc}$ and O$_{\rm hcp}$ is less (0.11~eV).

Comparing the average adsorption energies of the 
structures described above for coverage 0.50~ML  to 
the structures at same coverage,
but where oxygen occupies exclusively either on-surface 
($E_{\rm ad}$ = 2.92~eV for fcc oxygen) or sub-surface
sites ($E_{\rm ad}$ = 2.94~eV for O$_{\rm octa}$), 
it is found that the mixed structures,
$({\rm O}_{\rm fcc}/{\rm O}_{\rm tetra-I})_{\theta=0.50}$ (3.27~eV) and 
$({\rm O}_{\rm hcp}/{\rm O}_{\rm octa})_{\theta=0.50}$ (3.33~eV), are
significantly energetically  {\em more favorable}. 
The average adsorption energies of these mixed structures are,
however, not greater than that of on-surface adsorption at the
lower coverage of 0.25~ML.
Other geometries are also possible for coverage 0.50~ML, for example, 
pure on-surface
O involving hcp and fcc sites,
$({\rm O}_{\rm fcc}+{\rm O}_{\rm hcp})_{\theta=0.50}$, 
and pure sub-surface
O in different sites,
$({\rm O}_{\rm octa}+{\rm O}_{\rm tetra-II})_{\theta=0.50}$
and $({\rm O}_{\rm tetra-I}+{\rm O}_{\rm tetra-II})_{\theta=0.50}$,
where the two sub-surface
oxygen atoms are separated as far as possible from 
each other. All these geometries were tested and found to be energetically
unfavorable, as shown by their respective average adsorption energies,
namely,
2.41, 2.94 and 2.95~eV (cf. Tab.~\ref{tab:incorp}).
Altogether, our
calculations indicate 
that oxygen will enter the sub-surface region for coverages greater than
  0.25~ML.
We note that this coverage could be less, 
but to determine this would require the use of larger supercells; the
value of $\theta=0.25$~ML may therefore be regarded as an upper limit.
Clearly this is in contrast to O adsorption 
on Ru(0001) where O adsorption in sub-surface sites only
becomes energetically preferred after
completion of a full monolayer.~\cite{stampfl-oprl,apa}

Viewing oxygen as a typical electronegative atom, its function at the 
sub-surface site may be replaced by other electronegative species, 
like chlorine, which is actually used as a promoter in some reactions that 
silver catalyses.~\cite{sant87} It may be expected then that chlorine will 
also stabilize on-surface electronegative species. Also, the found enhanced 
stability of on-surface oxygen due to sub-surface O may also 
reflect the behavior of other similar 
species at the surface. This could explain the reported
stabilization of CO on Ag(100) which was attributed to
sub-surface oxygen.~\cite{rocc01}

\subsection{Atomic structure and electronic properties
\label{sec:eff-st-el}}

In the following we describe the main structural 
and electronic
properties of the energetically 
most favorable on-surface+sub-surface structures. Full details concerning 
the O-Ag bondlengths and metal interlayer spacing are given in 
Tab.~\ref{tab:sub1}. 
The most notable feature of the low energy structures is a characteristic 
linear O$_{\rm on}$-Ag-O$_{\rm sub}$ bonding arrangement of the uppermost 
atoms. In-line with their energetic preference, they also have shorter 
O-Ag bondlengths; for example, for 
$({\rm O}_{\rm fcc}/{\rm O}_{\rm tetra-I})_{\theta=0.50}$ the O$_{\rm fcc}$-Ag 
bondlength, $R_{\rm 1}^{\rm on}$, and the bondlength of sub-surface O to the 
first 
Ag layer, $R_{\rm 1}^{\rm sub}$, are 2.10 and 2.09~\AA\,, respectively, and for
$({\rm O}_{\rm hcp}/{\rm O}_{\rm octa})_{\theta=0.50}$,
the analogous values are 2.15 and 2.11~\AA\,, respectively.
These can be seen from inspection of
Tab.~\ref{tab:sub1} to be shorter than
the less favorable on-surface+sub-surface geometries.
They are also shorter compared to the corresponding
bondlengths of the pure on-surface and pure sub-surface structures,
e.g., 2.17~\AA\, for on-surface fcc oxygen, 2.12 for  
O$_{\rm tetra-I}$, and 2.24~\AA\, for O$_{\rm octa}$ at coverage 0.25~ML.
Compared to the values of $R_{\rm 1}^{\rm sub}$ (bondlengths of sub-surface
O to the top Ag layer), for the two most favorable structures,
the bondlengths of sub-surface oxygen to the {\em second} Ag layer (not shown here)
are pronouncedly longer, which indicates a stronger bonding of sub-surface O to 
the {\em surface} Ag layer, which is also bonded to a surface O atom.

In Tab.~\ref{tab:sub1} the work function changes with respect to the clean 
Ag(111) surface due to O adsorption  for the on-surface+sub-surface 
structures described above are listed. 
The values vary from 0.84 to 1.81~eV, and for the
energetically most favorable structures the values are,
$({\rm O}_{\rm fcc}/{\rm O}_{\rm tetra-I})_{\theta=0.50}$ 1.15~eV
and $({\rm O}_{\rm hcp}/{\rm O}_{\rm octa})_{\theta=0.50}$,
1.02~eV, respectively. 
For comparison, the change in work function for on-surface 
oxygen at coverage 0.25~ML
are 1.23~eV (fcc) and 1.31~eV (hcp),~\cite{wxli01} 
and the values of the work function for the on-surface
fcc and hcp sites at the same coverage of 0.50~ML are more
than twice as large as the structures involving
both on- and sub-surface oxygen.~\cite{wxli01}
Sub-surface O competes with on-surface
oxygen for the bonding charge of the surface Ag atoms,
as a consequence, the on-surface O species are slightly
{\em less negatively charged} compared to when there is no sub-surface O.
This can seen by comparing Fig.~\ref{fig:mix-0.50} (below)
with that of Fig.~4 in Ref.~\onlinecite{wxli01}, and is also indicated
by the lower energy position of the O-2$s$ levels for the
favorable on-surface+sub-surface structure as discussed  below.

\begin{figure}[t!]
\epsfxsize=0.45\textwidth \centerline{\epsfbox{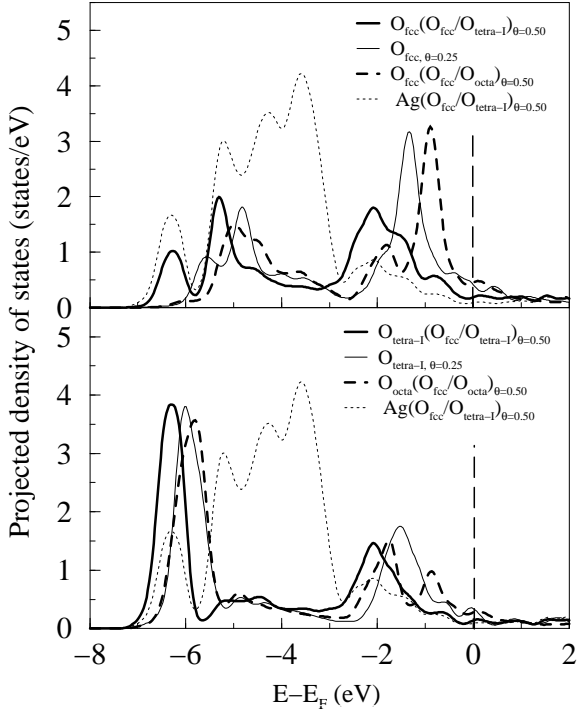}}
\caption{Total projected density of states (PDOS) for
the favorable
$({\rm O}_{\rm fcc}/{\rm O}_{\rm tetra-I})_{\theta=0.50}$
and unfavorable
$({\rm O}_{\rm fcc}/{\rm O}_{\rm octa})_{\theta=0.50}$
structures. PDOS for O$_{\rm fcc}$ (upper figure) and O$_{\rm tetra-I}$ 
(lower figure) are shown by thick solid lines, and the silver atoms shared 
by the two oxygen atoms are shown as dotted lines.  For comparison, 
O$_{\rm fcc}$ without the presence of sub-surface oxygen (thin solid line) 
and with sub-surface oxygen in the octa site (thick dashed line) are 
shown in the upper figure. Also for comparison (lower figure) are the 
PDOS for O$_{\rm tetra-I}$ without on-surface oxygen (shown as thin solid line)
as well as sub-surface O in the octa site of the unfavorable structure
(thick dashed line). The energy zero is the Fermi energy.
\label{fig:dos-fcc-t1-1}}
\end{figure}

\begin{figure}[t!]
\epsfxsize=0.35\textwidth \centerline{\epsfbox{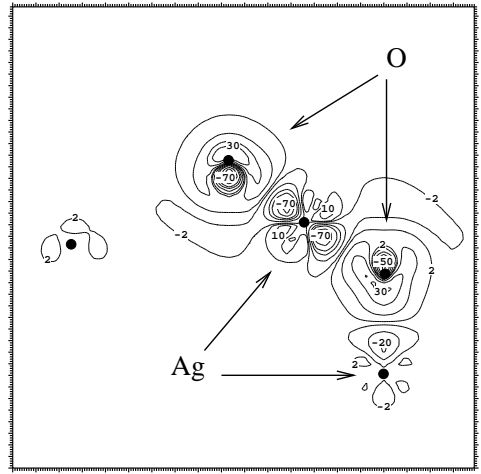}}
\caption{The difference electron density, 
\protect$n({\bf r})^{\Delta}$ 
(see the definition in Ref.~\protect\onlinecite{wxli01})
for $({\rm O}_{\rm fcc}/{\rm O}_{\rm tetra-I})_{\theta=0.50}$. 
The contour plane is in the [211] direction and is perpendicular to 
the
Ag(111) surface. The unit is  10$^{-3}$~Bohr$^{-3}$. The positions
of the O and Ag atoms are indicated by the arrows. }
\label{fig:mix-0.50}
\end{figure}

In Fig.~\ref{fig:dos-fcc-t1-1} we show the projected density of states for  
O$_{\rm fcc}$ and O$_{\rm tetra-I}$
atoms of the energetically
favorable $({\rm O}_{\rm fcc}/{\rm O}_{\rm tetra-I})_{\theta=0.50}$
structure. They are labeled
``${\rm O}_{\rm fcc}({\rm O}_{\rm fcc}/{\rm O}_{\rm tetra-I})_{\theta=0.50}$''
and "${\rm O}_{\rm tetra-I}({\rm O}_{\rm fcc}/{\rm O}_{\rm tetra-I})_{\theta=0.50}$ 
respectively, and are denoted by thick continuous lines in the upper and
lower parts of the figure.
We also show the PDOS for the first layer silver atoms which are
bonded to these two O atoms (dotted lines in Fig.~\ref{fig:dos-fcc-t1-1}).
For comparison we show the PDOS for the O$_{\rm fcc}$ and
O$_{\rm octa}$ atoms of the energetically {\em unfavorable}
$({\rm O}_{\rm fcc}/{\rm O}_{\rm octa})_{\theta=0.50}$ structure
(thick dashed curves in the upper and lower parts of the figure,
respectively), as well as the PDOS for the structures containing purely
O$_{\rm fcc}$ (i.e., without O$_{\rm tetra-I}$) and
purely O$_{\rm tetra-I}$ (i.e., without O$_{\rm fcc}$) atoms.

Considering first the results for 
$({\rm O}_{\rm fcc}/{\rm O}_{\rm tetra-I})_{\theta=0.50}$, 
it can be seen that the PDOS for O$_{\rm fcc}$ of 
this structure
is shifted to a lower energy compared to the pure on-surface
O$_{\rm fcc}$ species (compare the thick and thin continuous lines
respectively, in the upper plots).
The antibonding states of O$_{\rm fcc}$
of the $({\rm O}_{\rm fcc}/{\rm O}_{\rm tetra-I})_{\theta=0.50}$
structure are around 0.74~eV lower compared to the case of pure 
O$_{\rm fcc}$.  
A similar shift occurs for the O-2$s$ state. 
For O$_{\rm tetra-I}$ of the $({\rm O}_{\rm fcc}/{\rm O}_{\rm tetra-
I})_{\theta=0.50}$ 
structure, as compared to O$_{\rm tetra-I}$ of the
pure sub-surface O$_{\rm tetra-I}$ at 0.25~ML, 
it exhibits a similar behavior but the magnitude of the shift is less, 
i.e., around 0.56~eV for the valence PDOS and 0.45~eV for the 2$s$-state. 
Due to the mentioned shifts to lower energies, the DOS at the 
Fermi energy decrease for the
$({\rm O}_{\rm fcc}/{\rm O}_{\rm tetra-I})_{\theta=0.50}$
system  compared to that not containing sub-surface oxygen. This is
also the case for the {\em total} DOS.
The density of states
for the other energetically favorable structure, 
$({\rm O}_{\rm hcp}/{\rm O}_{\rm octa})_{\theta=0.50}$, (not shown)
are very similar to that of $({\rm O}_{\rm fcc}/{\rm O}_{\rm tetra-I})_{\theta=0.50}$ 
which is in-line with their similar
energetics. 

We now consider an {\em energetically unfavorable} configuration for
comparison, namely, 
$({\rm O}_{\rm fcc}/{\rm O}_{\rm octa})_{\theta=0.50}$, 
(where O$_{\rm octa}$ is 
in the 2nn set of O$_{\rm fcc}$, see Fig.~\ref{fig:schem} and 
Tab.~\ref{tab:sub1}). 
A different behavior to that described above is found: Here the PDOS for 
on-surface oxygen (O$_{\rm fcc}$) shifts towards {\em higher}
energies with the presence of sub-surface O, as does the O-2$s$ state 
(compare the thick dashed and thin continuous
lines in the upper plot of  Fig.~\ref{fig:dos-fcc-t1-1}).
Furthermore, we find that
all the energetically unfavorable structures exhibit
a higher total DOS at the Fermi level compared to the energetically
favorable ones.

To gain further insight into the bonding mechanism, we calculate the difference 
electron density distribution for
$({\rm O}_{\rm fcc}/{\rm O}_{\rm tetra-I})_{\theta=0.50}$, as shown in
Fig.~\ref{fig:mix-0.50}. A depletion of Ag-4$d_{\rm xz,yz}$ states of the
surface silver atoms bonded to the O$_{\rm fcc}$ and O$_{\rm tetra-I}$
oxygen atoms occurs, while there is an increase in electron density
on the O atoms.  The coupling between O$_{\rm tetra-I}$ and the silver atom
directly below in the second metal layer is comparably weak, as indicated by
the modest perturbation of electron density, and as also suggested by the
longer bondlength described above. It is the
O$_{\rm fcc}$-Ag-O$_{\rm tetra-I}$ linear coordination that allows a strong
hybridization of the Ag-4$d_{\rm xz,yz}$ orbital with the O$_{\rm fcc}$ and
O$_{\rm tetra-I}$ species, and also screens well the electrostatic interaction
between them. This is also the case for the 
$({\rm O}_{\rm hcp}/{\rm O}_{\rm octa})_{\theta=0.50}$ system
(not shown) which contains
the same linear coordination (i.e., O$_{\rm hcp}$-Ag-O$_{\rm octa}$) and weaker
coupling to the substrate underneath. 

The energetic preference for the fcc/tetra-I and hcp/octa (i.e.
a linear O-metal-O) arrangement is
found also at higher coverage. In particular,
we also considered  oxygen coverages of
2.00~ML. These structures involve
a full monolayer on the surface and full monolayer under the 
first Ag layer, where there
are a total of six possible configurations. We tested all of these and
as mentioned above, the energetically favorable structure is again 
that involving on-surface O in the hcp site and sub-surface O in
the octa site,$({\rm O}_{\rm hcp}/{\rm O}_{\rm octa})_{\theta=2.00}$ 
(by 0.07~eV), and the next most favorable structure is that
for on-surface O in the fcc sites and sub-surface O atoms in the tetra-I
sites, $({\rm O}_{\rm fcc}/{\rm O}_{\rm tetra-I})_{\theta=2.00}$.
The average adsorption energies of these structures, are however,
unstable with respect to gas phase O$_{2}$.
Such linear O-metal-O geometries have also been found 
to be the most energetically favorable of those considered
for other
transition metals, e.g. Ru~\cite{karsten} and Rh.~\cite{piro} 
As will be seen below, this is actually the primary building
block involved for the energetically favorable structures involving higher
oxygen concentrations as well as in silver(I)
oxide, Ag$_{2}$O. In addition, oxygen adsorption on Ag(110) with $(n\times1)$
reconstructions~\cite{tani} involve a similar O-Ag-O configuration.

\begin{figure}[b!]
\centerline{\psfig{figure=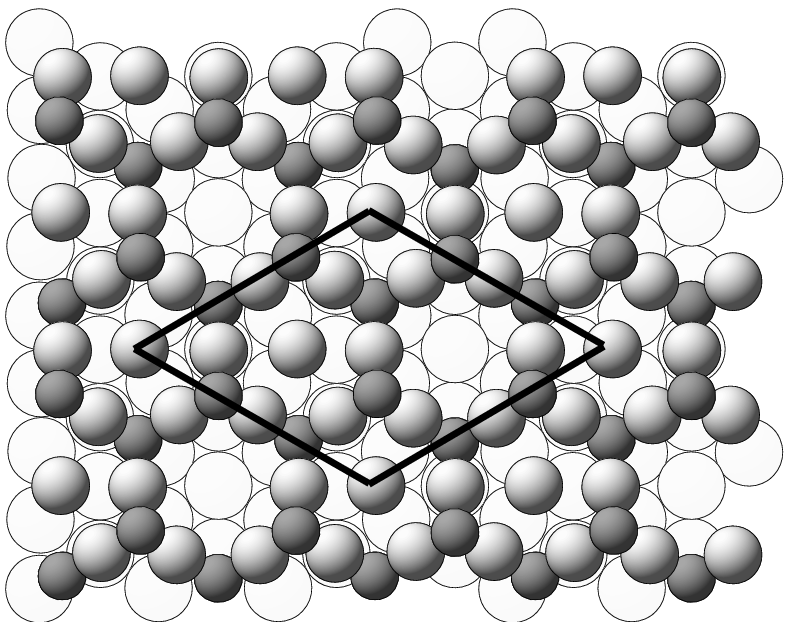,clip=,width=50mm}}
\caption{Sketch of the atomic structure proposed in 
the STM study of
Ref.~\protect\onlinecite{carl00} for the $(4\times4)$ phase.  The surface unit
cell is indicated. The oxygen atoms are represented by small dark circles, the
uppermost Ag atoms by grey circles, and the intact plane of Ag(111) atoms lying
below the `O-Ag-O' trilayer are represented as the open circles.}
\label{temp}
\end{figure}

\subsection{The $(4\times 4)$-O/Ag(111) phase}

The proposed atomic geometry for the $(4\times 4)$-O/Ag(111) 
phase~\cite{carl00,carl001} bears a very close resemblance 
to the identified low energy
$({\rm O}_{\rm fcc}/{\rm O}_{\rm tetra-I})_{\theta=0.50}$ structure.
The difference being that the surface Ag atoms of the
former structure
are {\em not} commensurate with the underlying $(1\times 1)$
surface unit cell, i.e. the upper O-Ag-O `tri-layer' is laterally
expanded to equal that of the (111) surface of bulk Ag$_{2}$O.
In addition, 
to avoid Ag atoms sitting in top sites directly on top of the underlying
Ag atoms, these Ag atoms are missing, i.e.
one per $(4\times4)$ cell. This results
in a stoichiometry of the tri-layer of Ag$_{1.83}$O with
a corresponding oxygen coverage of 
0.375~ML (see Fig.~\ref{temp}).
We calculated the average O adsorption energy of this structure 
(including full atomic optimization) and
find that it has a low energy
(see Fig.~\ref{energy2}). 
In obtaining this value, we have used the chemical potential
of Ag in bulk to take into account the energy of the Ag atoms in
the mixed tri-layer.
We also considered a `stoichiometric structure' i.e., corresponding
to Ag$_{2}$O within the tri-layer in which the
Ag atoms sitting on top of underlying substrate atoms are
present and found, in 
agreement with other recent calculations of these two geometries,~\cite{wien}
that it is less favorable by 0.7~eV per cell.

\section{Formation of a surface oxide}

\subsection{Accumulation of oxygen in the sub-surface region}

In order to gain insight into the stability of 
structures with higher concentrations of oxygen, we 
performed calculations for many different geometries. Since
there are no experimental results to provide guidance, 
and no additional ordered phases observed, we use $(2\times 2)$
surface cells and systematically investigate the many possible atomic
configurations at the given coverages. Such a study would
be unfeasible with the larger $(4\times 4)$ cells, and given
the very similar local atomic geometry mentioned above 
of the proposed $(4\times 4)$ structure and our identified
low energy structure at 0.5~ML (and similar electronic
structure), we view this as an appropriate strategy.

\begin{figure}[b!]
\epsfxsize=0.45\textwidth \centerline{\epsfbox{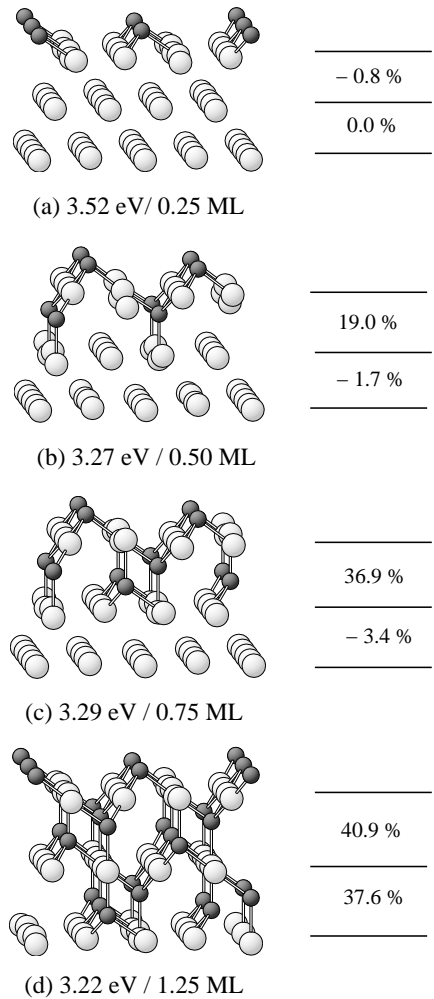}}
\caption{Atomic geometry of energetically
favorable structures for increasing
oxygen concentrations:
   (a)  on-surface fcc oxygen, (b) fcc oxygen plus
sub-surface oxygen in the tetra-I site, (c) as for (b) but with additional
oxygen in the 
  tetra-II site, (d) 
  as for (c) but with additional oxygen under the second Ag layer in the
  tetra-I and tetra-II sites.
The average adsorption energy with
  respect to the clean Ag(111) substrate and free oxygen atoms, as well
  as the corresponding coverage, are 
  given at the bottom of each figure. The relative variation of the first and
  second interlayer spacings with respect to the bulk value 
  is also given to the right of the figures. Large pale grey and
small dark circles
  represent silver and oxygen atoms, respectively.}
\label{fig:oxide}
\end{figure}

We first investigate
various geometries involving 0.75~ML.  
Considering the addition of another O atom to the energetically
favorable systems with total 
O coverage 0.50~ML,
our results show that oxygen will
occupy sub-surface sites rather than on-surface sites to avoid the strong
dipole-dipole repulsion. As seen in Tab.~\ref{tab:incorp}, for the 
energetically favorable configuration  described above for 0.50~ML of oxygen,
$({\rm O}_{\rm hcp}/{\rm O}_{\rm octa})_{\theta=0.50}$, the addition of an 
O atom to the on-surface hcp site yields an average binding energy of 2.97~eV,
but it is 3.25~eV if the additional O atom occupies
the sub-surface octa site below the first Ag layer,
$({\rm O}_{\rm hcp}/{\rm O}_{\rm octa})_{\theta=0.75}$.
A similar result is found for the other energetically favorable structure at 
0.50~ML, $({\rm O}_{\rm fcc}/{\rm O}_{\rm tetra-I})_{\theta=0.50}$, in        
that it is energetically favorable for the additional oxygen atom to
occupy a sub-surface site, namely the tetra-I site,
$({\rm O}_{\rm fcc}/{\rm O}_{\rm tetra-I})_{\theta=0.75}$,
(3.06~eV), instead of an on-surface fcc-hollow site (2.90~eV). 
The reason that the average binding energy of
$({\rm O}_{\rm hcp}/{\rm O}_{\rm octa})_{\theta=0.75}$ with 0.50~ML
octa oxygen is 0.19~eV more favorable than 
$({\rm O}_{\rm fcc}/{\rm O}_{\rm tetra-I})_{\theta=0.75}$ with 0.50~ML
tetra-I oxygen, is mainly due to the difference in energy of the pure sub-
surface O$_{\rm octa}$ and the  pure sub-surface O$_{\rm tetra-I}$ structures 
at coverage 0.50, which as shown in Tab.~\ref{tab:sub}, is 2.57~eV for the
latter and 2.94~eV for the former. 

We also tested if it is energetically more favorable for the
two sub-surface O species to occupy
different types of sub-surface sites.
For instance, as shown in Tab.~\ref{tab:incorp},
(O$_{\rm octa}$+O$_{\rm tetra-II}$)$_{\theta=0.50}$ and 
(O$_{\rm tetra-I}$+O$_{\rm tetra-II}$)$_{\theta=0.50}$, have average adsorption 
energies of 2.94 and 2.95~eV, respectively, very close to O$_{\rm octa}$ at 
same coverage (2.94~eV). 
We find that the energetically most favorable structures at 
coverage 0.75~ML do involve
sub-surface O atoms in different sites, in addition to
that where they both occupy the octa site, namely,
$({\rm O}_{\rm hcp}/{\rm O}_{\rm octa}+{\rm O}_{\rm tetra-II})_{\theta=0.75}$ 
with average adsorption energy 3.26~eV, 
$({\rm O}_{\rm fcc}/{\rm O}_{\rm tetra-I}+{\rm O}_{\rm
tetra-II})_{\theta=0.75}$ with average binding energy 3.29~eV, as well
as, $({\rm O}_{\rm hcp}/{\rm O}_{\rm octa})_{\theta=0.75}$ (3.25~eV).

Interestingly, the slightly more favorable one,
$({\rm O}_{\rm fcc}/{\rm O}_{\rm tetra-I}+{\rm O}_{\rm tetra-II})_{\theta=0.75}$ 
(depicted in Fig.~\ref{fig:oxide}c), is similar in structure to the bulk silver 
oxide, Ag$_{\rm 2}$O, where both sub-surface oxygen atoms are located at
the center of a silver tetrahedron, and the silver atoms shared by the O atoms 
are in a linear chain. Within the accuracy of our calculations, however,
these three structures are essentially degenerate.

\begin{table}[t!]
\caption{Average adsorption energy (per oxygen atom), 
$E_{\rm ad}$, and work function change, $\Delta\Phi$, for oxygen in 
various sites and for various coverages. 
$\theta_{\rm sub}$, $\theta_{\rm on}$ and $\theta_{\rm total}$ 
represent the oxygen coverage in the sub-surface and on-surface 
regions, and the total coverage, respectively. 
The energy unit is eV. }
\label{tab:incorp}
\begin{tabular}{l|cccc}
    & $\theta_{\rm on} $ 
    & $\theta_{\rm sub}$ 
    & $E_{\rm ad}$ 
    & $\Delta\Phi$ \\ \hline 
$ \theta_{\rm total}=0.25$ 
                  &            &       &      &       \\ 
fcc               &  0.25      & -     & 3.52 & 1.23 \\ 
octa              & -          &0.25   &2.85  &0.09 \\ \hline
$ \theta_{\rm total}=0.50$ 
                  &            &       &      & \\ 
fcc               & 0.50       &-      &2.92  &2.22 \\ 
hcp               & 0.50       &-      &2.83  &2.36 \\ 
fcc+hcp           & 0.50       &-      &2.41  &2.53 \\ 
octa              & -          &0.50   &2.94  &0.09    \\ 
octa+tetra-II     & -          &0.50   &2.94  &0.02  \\
tetra-I+tetra-II  & -          &0.50   &2.95  &0.27  \\
octa+tetra-I      & -          &0.50   &2.78  &0.46 \\
{\bf fcc/tetra-I} & {\bf 0.25} &{\bf 0.25}    & {\bf 3.27} 
                  &{\bf 1.15} \\  
hcp/octa          & 0.25       & 0.25  &3.33  & 1.02 \\\hline  
$ \theta_{\rm total}=0.75$                   
                  &            &       &      & \\ 
hcp/octa          & 0.50       &0.25   &2.97  & 2.22 \\ 
hcp/octa          & 0.25       &0.50   &3.25  & 1.42 \\ 
hcp/octa+tetra-II    
                  & 0.25       & 0.50  & 3.26 & 0.85 \\
fcc/tetra-I       & 0.50       &0.25   & 2.90 &2.19 \\ 
fcc/tetra-I       & 0.25       &0.50   &3.06  &1.47 \\ 
{\bf fcc/tetra-I+tetra-II} 
                  &{\bf 0.25}  &{\bf 0.50} &{\bf 3.29}  
                  &{\bf 0.98} \\ \hline  
$ \theta_{\rm total}\ge1.00$                   
                  &            &        &      & \\ 
hcp/octa          & 0.25       &0.75    &3.09  &1.63 \\ 
hcp/octa          & 0.25       &1.00    &2.89  &1.58 \\ 
{\bf fcc/tetra-I+tetra-II} 
                  &            &        &      & \\                    
{\bf ~~~~/tetra-I}     
                  &{\bf 0.25}  &{\bf 0.75} &{\bf 3.13}  
                  &{\bf 0.66} \\ 
{\bf fcc/tetra-I+tetra-II}
                  &            &         &    & \\ 
{\bf ~~~~/tetra-I+tetra-II} 
                  &{\bf 0.25}  &{\bf 1.00}   &{\bf 3.22}  
                  & {\bf 1.02} \\ \hline
{\bf fcc/tetra-I+tetra-II}
                  &            &       &     & \\ 
{\bf ~.../tetra-I+tetra-II} 
                  &{\bf 0.25}  &{\bf 2.00}   &{\bf 3.21}  
                  & {\bf 1.17} \\ 
\end{tabular}
\end{table}

\subsection{Higher O concentrations and formation of an oxide-like film}

We now consider the incorporation of higher concentrations of atomic
oxygen. For a coverage
of 0.75~ML the favored structures have 0.25~ML oxygen
on the surface and 0.50~ML between the first and second Ag layers.
We will now determine the energetic preference for adding one more O
atom and then subsequently a second one, resulting
in total coverages of 1.00~ML and 1.25~ML, respectively.
In particular, we will investigate whether oxygen will
prefer to occupy the sub-surface region under the first metal layer
along with the two oxygen atoms already there, or if the O atoms will
reside deeper in the ``bulk'' region, i.e., under the second metal layer. 

The results are shown in the lower part of Tab.~\ref{tab:incorp}.  For
additional oxygen occupation under the first Ag layer,
as an example, we show for the hcp/octa system the result
for sub-surface coverages of 0.75 and 1.00~ML with O in the octa site.
The average adsorption energies are 3.09 and 2.89~eV, respectively.
However, if these additional oxygen atoms occupy instead sites under the
{\em second} Ag layer, in particular, according to the identified 
preference for a linear O-Ag-O bonding arrangement, then 
the average adsorption energies are 3.13 and 3.22~eV,
respectively, which  are clearly energetically
more favorable compared to 
additional oxygen occupation under the first Ag layer.
The latter structure is  shown in Fig.~\ref{fig:oxide}d, where it can be
seen, in comparison with \ref{fig:oxide}b and \ref{fig:oxide}c, that
with continuous O incorporation, the same structure can keep
forming. Actually, the average binding energy of
3.22 eV at coverage 1.25~ML, has
already converged to the value
of the structure at the higher coverage
of 2.25~ML (3.21~eV), as shown at the end of Tab.~\ref{tab:incorp}.
This value is also very close to the calculated heat of
formation for {\em bulk} Ag$_2$O, which is
3.23~eV (also referred to the free O atom).~\cite{wxli02}

\begin{figure}[t]
\epsfxsize=0.45\textwidth \centerline{\epsfbox{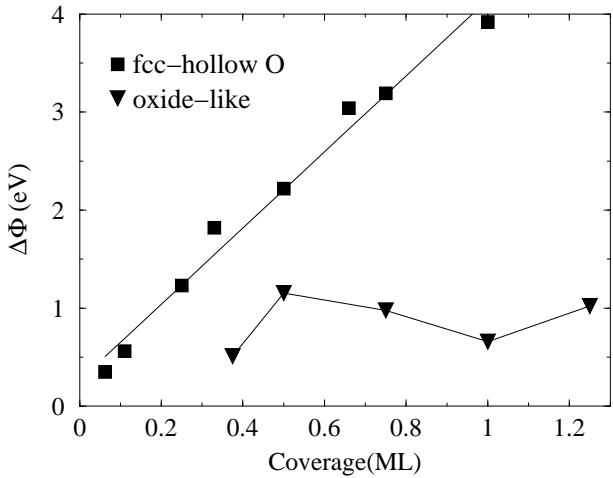}}
\caption{Work function change for on-surface O adsorption in the fcc site 
(squares)
(from Ref.~\protect\onlinecite{wxli01}) and of the 
$(4\times 4)$ and the oxide-like structures 
with higher oxygen concentrations
 (triangles).}
\label{fig:workfunction}
\end{figure}

In Fig.~\ref{fig:workfunction} the work function change of the identified energetically
favorable structures involving on- and sub-surface oxygen,
as well as that for the $(4\times4)$ structure are shown, where
they are compared to the values for pure on-surface adsorption. It can be
seen that they are significantly lower than for on-surface adsorption and vary
between
1.1 to 0.5 eV.
Interestingly these values are close to that of $\sim$1.0 eV as reported experimentally
for Ag(111) when
subjected to high oxygen pressures and high temperature.~\cite{bao-prb}
The value for the $(4\times4)$ structure is also in very good agreement with
the value of 0.55~eV reported by Bare {\em et al.} ~\cite{bare95}

\begin{figure}[t!]
\epsfxsize=0.45\textwidth \centerline{\epsfbox{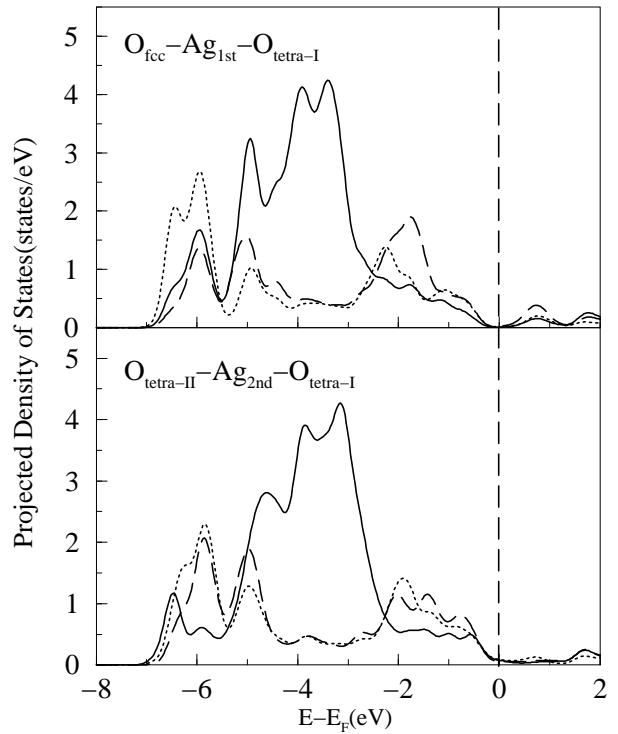}}
\caption{Total projected density of states for the oxide-like structure 
(O$_{\rm fcc}$/O$_{\rm tetra_I}$+O$_{\rm tetra-II}$/O$_{\rm tetra_I}$+O$_{\rm
  tetra-II})_{\theta=1.25}$.
 The upper figure shows the PDOS for 
O$_{\rm fcc}$ (dashed line), the uppermost Ag atoms bonded to both
O atoms (solid line)
and O$_{\rm tetra-I}$ (dotted line) under the first layer. The lower   
figure
shows the PDOS for O$_{\rm tetra-II}$ (dashed line)
under the first metal layer, the shared Ag atom 
(solid line) and O$_{\rm tetra-I}$
(dotted line) under the second metal layer.
The Fermi energy is indicated by the vertical dashed line.
\label{fig:dos-oxide-like}}
\end{figure}

The identified energetically favorable structures with high O concentrations 
{\sl e.g.} 1.25~ML or 2.25~ML are like silver(I) oxide Ag$_2$O(111)
but under a compressive lateral
strain of 18$\%$.
The character of the oxide is already present for 
coverage 0.75~ML: In addition to the same
local atomic coordination, the vibration of sub-surface oxygen, 
O$_{\rm tetra-I}$, in 
(O$_{\rm fcc}$/O$_{\rm tetra-I}$+O$_{\rm tetra-II}$)$_{\theta=0.75}$ 
is calculated to be 526~cm$^{-1}$ which is
very similar to the experimental value of
545~cm$^{-1}$ for oxygen in bulk Ag$_2$O.~\cite{pett94}
The calculated valence band for the oxide-like structure 
(O$_{\rm fcc}$/O$_{\rm tetra-I}$+O$_{\rm tetra-II}$/O$_{\rm tetra-I}$
+O$_{\rm tetra-II}$)$_{\theta=1.25}$, shown in
Fig.~\ref{fig:dos-oxide-like}, also bears a close resemblance
to that of bulk Ag$_2$O. The upper panel shows the result for the
uppermost O$_{\rm fcc}$-Ag-O$_{\rm tetra-I}$  atoms, and the lower one shows
the result for O$_{\rm tetra-II}$ under the first metal layer and 
O$_{\rm tetra-I}$ under the second metal layer, as well as the silver atom  
in the second layer to which they commonly bond. 
Compared to the clean surface, or lower coverage structures, e.g.,
(O$_{\rm fcc}$/O$_{\rm tetra-I}$)$_{\theta=0.50}$ as shown in 
Fig.~\ref{fig:dos-fcc-t1-1}, the silver 4$d$-band has considerably narrowed 
due to 
the significant increase of the spacing between the metal layers.
Also, two extended shoulders appear either side of it.  These
features are
very similar to those of the the oxide surface under zero lateral
strain, which will be reported in detail elsewhere.~\cite{wxli02} 
We calculated the removal energies of the
on-surface oxygen of the favorable structures 
listed in Tab.~III.
The resulting energies are very similar to what we found for the
favorable 0.50~ML on-surface+sub-surface structures discussed
above in Sec.~IV, namely, they are all in the range of
3.90$\sim$4.01~eV.  These bondstrengths are actually the
strongest we have found for this system and such species
could be important for understanding how silver functions
as an efficient oxidation catalyst. 

\subsubsection{Relative stability of the oxide-like structures}
It has been seen above that the average O
adsorption energies of the energetically favorable structures
at $\theta$=0.50 and 0.75~ML are slightly greater than 
the heat of formation of
the bulk oxide referred to atomic oxygen
(compare 3.27 and 3.29~eV to 3.23~eV). For the 
higher concentration structures the values are however extremely
close (compare 3.22 and 3.21 to 3.23~eV).
This is very interesting since the oxide-like structures, although
similar in geometry to ideal Ag$_{2}$O(111)  
surfaces, are laterally compressed by a large amount.
To investigate why the energies are so close,
e.g., whether the underlying
Ag(111) surface plays a role, or if the distortion actually has little
effect on the energetics,
we study the relative stability of the oxide-like structures in comparison
to  true Ag$_{2}$O(111) structures. 
In particular, we calculate free-standing
Ag$_{2}$O(111) layer structures under zero strain and under
a compressive strain of 18$\%$,
which corresponds to the strain of our identified oxide-like structures
due to being commensurate with the underlying Ag(111) surface. 

For a thickness involving
5 metal layers, it is found that the true oxide is
energetically {\em less} favorable by 0.05~eV per oxygen atom than 
the oxide film under the strain.
For thicker films corresponding to
7 metal layers, the energy
difference between two structures decreases to just 0.005~eV per oxygen 
atom, where the
deformed oxide film is still preferred. Only for 
thicker structures involving 9 metal layers
does the true oxide film (zero strain)
become energetically favorable (by
0.02~eV per oxygen atom).
This could be taken to indicate that a transition to the natural oxide surface
will only happen at or beyond a critical thickness corresponding to 
about nine metal layers. The transition may not
occur immediately 
because the energy gain is still very small and it may
take more layers in order for sufficient strain 
energy to build-up.
Also, the interaction between the oxide structures and 
the metal substrate, which was 
not considered, may hinder the transition.
That the structures with the smaller lateral dimensions
are not notably less favorable than the true
oxide structures
can be understood in that
there is a significant expansion normal to the surface, as shown in
Fig.~\ref{fig:oxide}. These energetics are in contrast to what was
found for the O/Ru(0001) system,~\cite{karsten} where from
similar calculations it was found that
RuO$_{2}$(110) layer structures become energetically more favorable
compared to layers of the 
identified on-surface+sub-surface structures 
with the same stoichiometry, after a thickness equivalent to only two
or more metal layers. In this case the lateral deviation between
the natural RuO$_{2}$(110) oxide and the 
on-surface+sub-surface structure
was equal to 15~\% and 35~\% in the $y$ and $x$ directions, respectively,
where the natural oxide is more expanded.

\subsubsection{Adsorption of an ozone-like species}
\begin{figure}[t!]
\epsfxsize=0.40\textwidth \centerline{\epsfbox{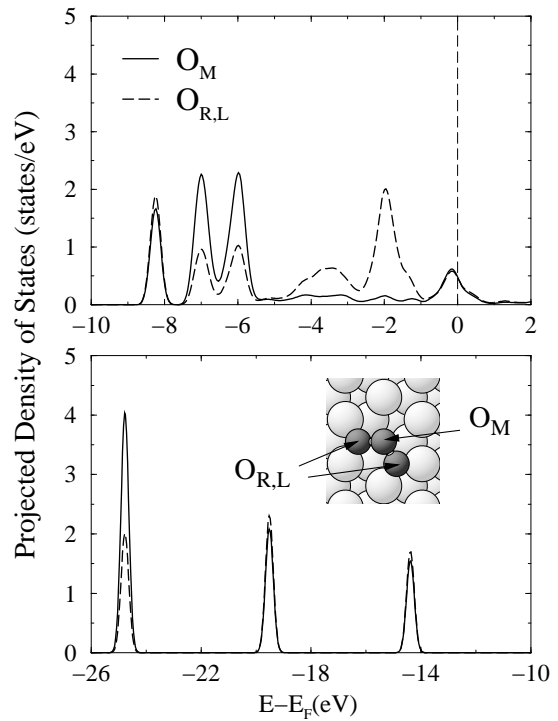}}
\caption{
Total projected oxygen density of states for the ozone-like
species adsorbed at an Ag surface vacancy.
The upper panel shows the valence band energy region and the bottom panel, 
the lower energy region. ``O$_{\rm M}$'' indicates the center
O atom of the ozone-like species and the ``O$_{\rm L/R}$'' indicate
the left and right O atoms of the ozone-like species that are bonded
to two surface Ag atoms.}
\label{dos-ozone}
\end{figure}

\begin{figure}[t!]
\epsfxsize=0.45\textwidth \centerline{\epsfbox{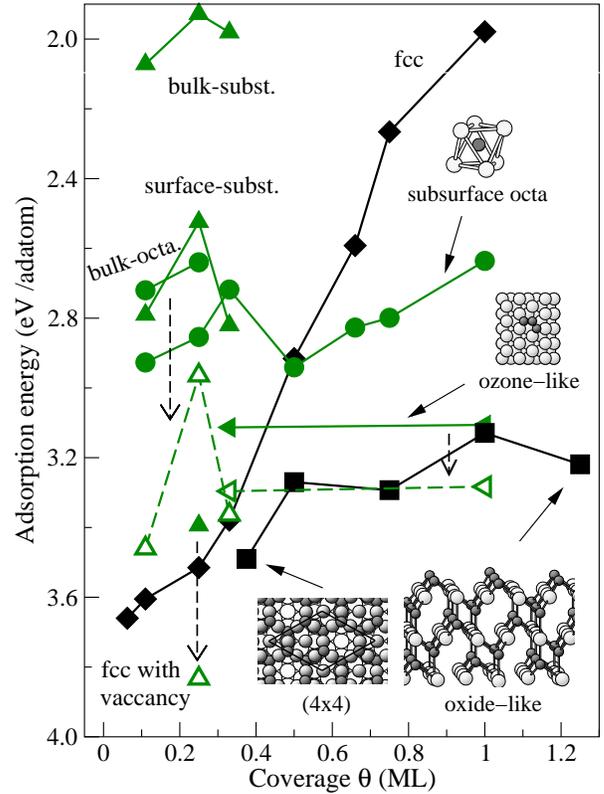}}
\caption{Average adsorption energy (with respect
to atomic O) versus coverage for the different
structures considered:
pure on-surface oxygen in fcc-hollow sites (diamonds), 
  surface- and bulk-substitutional adsorption (triangles),
 pure sub-surface oxygen in octahedral
  sites under the first (subsurface-octa.)
  and second (bulk-octa.) Ag layers (circles), an ozone-like molecule
  adsorbed at a surface Ag vacancy 
(left-pointing triangles), and surface 
oxide-like structures (squares), as 
  illustrated in Fig.~\ref{fig:oxide}. 
  The vertical downward pointing arrows directed at dashed lines with
  open symbols, indicate the corresponding 
{\em binding energies} of the structures involving surface Ag vacancies, i.e. the
vacancy formation energy, $E_{f}^{\rm vac}$ (see
Ref.~\protect\onlinecite{wxli01} for the definition), 
 is not taken into account. 
 The adsorption energy of O in the fcc site (triangle) with
 a neighboring surface Ag vacancy is also shown for $\theta=0.25$~ML,
 as well as the corresponding binding energy.}
\label{energy2}
\end{figure}

Another issue of debate in the literature is the origin of several
electronic states observed lying in the energy region below
the Ag-$4d$ band and above the O-2$s$ semi-core level.
On one hand
it has been proposed that hydroxl groups are responsible~\cite{bukh011}
and on the other, an associatively adsorbed ozone-like species
at a surface Ag vacancy has been proposed.~\cite{boron}
We investigated the latter species, the atomic 
geometry of which is shown in the insets of 
Figs.~\ref{dos-ozone} and \ref{energy2}.
We find, in agreement with Ref.~\onlinecite{boron}, that
the molecular ozone-like species is energetically preferred 
compared to three separate O atoms adsorbed at the vacancy edges.
We considered two vacancy 
concentrations for the adsorption of the ozone-like
species, namely, 0.11 and 0.33~ML, corresponding to
O coverages of 0.33 and 1.0~ML, respectively.
The atomic geometry is found to be similar to 
the ozone molecule in the gas phase, namely, when adsorbed on the
surface the calculated bondlength between the O atoms is 1.41~\AA\,
and the angle is 118 degrees, while for the free molecule the
respective values are 1.27~\AA\, and 117 degrees.
The longer bondlength between the O atoms when
adsorbed on the surface indicates that the inter O-O bonding 
is weakened.
The adsorption energy is very similar in each case, indicating
that they interact little with each other. Compared to the oxide-like
structures they are energetically unfavorable. However, if vacancies
were provided `for free' (i.e. not taking into account the
vacancy formation energy), 
then such a species
is energetically favorable for $\theta=1.0$~ML (see Fig.~\ref{energy2}).
We expect, however, that 
the concentrations of vacancies will be rather low and this
species will not be present in large concentrations.

With regard to the electronic structure, we
find that there are indeed a number of electronic
states in the energy regime below the bottom Ag-4$d$ band as can be seen from
Fig.~\ref{dos-ozone}, which shows
the  projected density of states on the O atoms.
In particular, features are seen at energies of $\sim$2.0, 3.5, 6.0, 7.0,
8.5, 14.4, 19.5, 24.8~eV below $E_{\rm F}$.
Reported experimental features observed on polycrystalline 
silver are  at 3.5, 9.2, and 11.2~eV below $E_{\rm F}$,
which were thought to be due to hydroxls,~\cite{bukh011}
and features observed for the  O/Ag(110) system are at
3.3, 7.0, 9.5, 12.7, 17.3~eV, which were attribued to the associatively
adsorbed ozone-like species.~\cite{boron}
It can be appreciated that there is
some overlap of these various energies 
but it is not possible to draw any solid
conclusions on the basis of such comparisons.

In order to visually summarize the main energetics of our results, we show
in Fig.~\ref{energy2} the average
adsorption of oxygen at Ag(111) versus coverage. This
also includes our earlier results for pure on-surface and surface-substitutional
adsorption~\cite{wxli01} 
as well as for investigations into  bulk substitutional adsorption.
There is one point in this figure not discussed in the present or
previous (Ref.~\onlinecite{wxli01}) paper so far and that is for 
the adsorption of an oxygen atom on the surface
in the fcc site
near a pre-existing vacancy with coverage 0.25~ML.
The adsorption energy is seen to be less favorable than on the
ideal (111) terrace, but if such vacancies were provided ``for free''
(i.e. without paying the vacancy formation energy cost) it can
be seen that their binding to the surface
is considerably stronger than on the ideal (111)
terrace, compare 3.83~eV to 3.52~eV. 
We expect a similarly stronger binding of O at other under-coordinated
metal sites such as at the edge of steps or kinks, as has been found
for other O/transition-metal systems.
Considering Fig.~\ref{energy2}, at a glance, these results predict that
at low coverage, O prefers to adsorb on the surface, but with increasing
coverage the thin $(4\times 4)$ surface-oxide is favored.
For higher concentrations, thicker oxide-like structures similar to the (111)
surface of Ag$_{2}$O are preferred which have a coverage of
0.25~ML on the surface and 0.50 ML between each Ag layer. 
These energetics relate well to the general behavior found experimentally 
in the early work of  
Czanderna,~\cite{czan77} whose study shows that the isosteric heat of 
adsorption 
of oxygen on silver powder 
decreases from 0.91 to 0.37 eV per oxygen atom
(with respect to 1/2 the binding energy of O$_{2}$), and then remains
constant at 0.40$\pm$0.02 eV per oxygen atom for
coverages $\theta$=0.33 to about 0.90~ML, which was attributed to the growth
of an oxide layer.

\subsection{Thermodynamic and kinetic considerations}
  
For a metal in contact with an oxygen atmosphere, in thermal equilibrium,
the bulk oxide will form when  the oxygen chemical potential, $\mu_{\rm O}$,
satisfies the equation,
\begin{eqnarray}
\mu_{\rm O} + 2 \mu_{\rm Ag(bulk)} = \mu_{\rm Ag_{2}O}
\quad ,
\end{eqnarray}
where $\mu_{\rm Ag(bulk)}$ and $\mu_{\rm Ag_{2}O}$ are the chemical potentials
of an Ag atom in bulk and of the bulk oxide, Ag$_{2}$O, respectively.
This  equation can be expressed in terms of the total energies of the
respective systems, i.e.,
\begin{eqnarray}
E_{\rm O} + 2 E_{\rm Ag} = E_{\rm Ag_{2}O}
\quad .
\end{eqnarray}
Using the above equation, and referring the energy to that of
a free O atom, we obtain
for the O chemical potential $\mu_{\rm O}=3.23$~eV. This is just the
heat of formation of Ag$_{2}$O, $\Delta H_{\rm oxide}$,
but referred to a free O atom instead of to 1/2O$_{2}$.

Concerning the transition from chemisorption to bulk oxide
formation,
it has been proposed that the critical coverage, $\theta_c$, at
which the transition occurs is thermodynamically,
rather than kinetically, determined. 
Below $\theta_c$, the
differential heat of adsorption of chemisorbed oxygen is higher than
the heat of formation of the oxide, but above $\theta_c$ 
it becomes lower.
The differential heat of adsorption at a given
coverage, $H^{\rm diff}_{\rm ads}(\theta)$,
can be thought of as the energy required to remove a single
O atom on the surface
at that coverage, or equivalently, the energy gain on adsorption of
a single O atom to result in that coverage 
(i.e., with the presence of other pre-adsorbed O atoms).
The criteria for the critical coverage beyond which
bulk oxide formation is expected to occur is,
\begin{eqnarray}
\Delta H_{\rm oxide} &=& H^{\rm diff}_{\rm ads}(\theta_{\rm c}) =\left .
\frac{d (\theta\times E_{\rm ads}(\theta))}{d \theta} \right | _{\theta_{\rm
  c}} \nonumber \\ 
&=& E_{\rm ads}(\theta_{\rm c})+\theta_{\rm c}\times 
\left. \frac{d E_{\rm ads}(\theta)} {d \theta} \right|_{\theta_{\rm c}}
\quad .
\end{eqnarray}
Applying this to the O/Ag(111) system, using the on-surface adsorption
energies from our previous paper,~\cite{wxli01} 
we obtain a critical coverage $\theta_c$ of
0.24~ML. This value correlates very closely to 
the coverage of 0.25~ML found
from our extensive DFT calculations
when oxygen starts to penetrate into the sub-surface region, suggesting
onset of oxide formation and occupation of sub-surface sites are linked.
Similar correlations for O at other transition 
metal surfaces, namely, Ru(0001), Rh(111), and Pd(111) occur,
where it is found that the critical
coverage is higher for the ``harder'' metals towards the left of the TM 
series (i.e., Ru).~\cite{mira01}

In contrast to our theoretical results which predict the formation of 
oxide-like films, to our knowledge, experimental studies employing
high pressures and elevated temperatures have only reported the
thin $(4\times4)$ surface-oxide
to date.~\cite{carl00,carl001,rovi74,camp85,bare95} 
Formation of an oxide 
requires firstly sufficient numbers of dissociated oxygen atoms, 
and secondly, it requires 
significant mass transport (e.g., diffusion of Ag atoms or O atoms)
which can only take place at elevated
temperatures. 
Given that the heat of formation of bulk silver oxide is low, we expect
that formation of thicker oxide-like layers may be prevented
either because the temperature at which they likely decompose is not high enough
for sufficient atomic rearrangements to occur, or that so far there
have been insufficient concentrations of oxygen atoms  
delivered to the surface, indeed 
the (111) surface of silver has an extremely low
dissociative sticking coefficient for O$_{2}$, 10$^{-6}$. 
Possibly with the use of atomic oxygen or ozone (which readily
dissociates at the surface) and low temperatures,
thicker oxide films
could be formed experimentally.
This has recently been done for studies
on the oxidation of the Pt(111) surface~\cite{saliba}
as well as
for polycrystalline silver when exposed
at room temperature to
a mixture of oxygen and ozone, where thick films have recently been
reported.~\cite{thick-oxide-layer}

In order to gain more insight into the behavior of the O/Ag system as an
oxidation catalyst and to identify possibly
active O species, it is important to be
able to take into account the effect of temperature and gas pressure.
In particular, the catalytic reactions of ethylene epoxidation and the partial 
oxidation of methanol take place at  two significantly different temperature
ranges (500-600~K and 700-900~K, respectively) and both occur at high
(atmospheric) pressure. Our investigations 
in this direction are in progress.

\section{Conclusions}

In the present work, we have systematically investigated pure sub-surface
oxygen and structures involving both oxygen adsorbed on the
surface and sub-surface oxygen, as well as 
oxide-like structures at the Ag(111) surface through density functional theory
calculations. Compared to pure on-surface oxygen adsorption, which exhibits a
strong decrease in adsorption energy with increasing coverage, pure sub-surface 
adsorption exhibits a weak dependence on the coverage, where
the octahedral site is favored for all of the investigated coverage 
range of 0.11 to 1~ML. 
Comparing the adsorption energies of
pure on-surface O and pure sub-surface
O, we find the former is energetically favored over the latter
for coverages up to around 1/2 a monolayer, whereafter
the pure sub-surface structures are preferred.

The energetically favorable structures involving both on-surface
and sub-surface oxygen have a lower density of states at the Fermi 
energy and involve less
ionic (less negatively charged) O atoms as compared
to the energetically unfavorable structures.
As quantified by the concept of a ``removal'' energy,
the presence of sub-surface oxygen modifies the on-surface oxygen-silver
bond significantly. Depending on the adsorption site,
it can {\em either stabilize or destabilize} on-surface oxygen, and vice versa. 
The energetically most favorable structures, 
however, {\em stabilize} on-surface oxygen. 

On the  basis of the energetics of all the calculated structures
we find the following scenario:
At low coverages 
oxygen prefers to stay on the surface in
the fcc site. With
increasing coverage a thin $(4\times 4)$ surface-oxide is
energetically favorable, and for higher concentrations,
the calculations predict
the formation of oxide-like structures
very similar to Ag$_{2}$O(111).
We propose that with the use of atomic oxygen or
ozone and low temperatures, such thicker oxide-like films
may be observed experimentally.

Having thoroughly studied the energetics, and atomic and electronic properties 
of the O/Ag(111) system,
as is crucial for gaining further understanding into the 
function of silver   
as an oxidation catalyst, the next considerations must 
address the effect 
of elevated temperatures and high pressures, since these are the conditions under 
which real catalysis takes place. This will be the subject of our following 
publications.

\end{document}